\documentclass[preprint,aps,prd,superscriptaddress,nofootinbib,showpacs]{revtex4}
\usepackage{epsfig}
\usepackage{graphicx}
\usepackage{dcolumn}
\usepackage{bm}

\date{}
\begin{document}


\newcommand{\D}{\displaystyle}
\newcommand{\T}{\textstyle}
\newcommand{\mc}{\multicolumn}
\newcommand{\bce}{\begin{center}}
\newcommand{\ece}{\end{center}}
\newcommand{\beq}{\begin{equation}}
\newcommand{\eeq}{\end{equation}}
\newcommand{\bea}{\begin{eqnarray}}

\newcommand{\eea}{\end{eqnarray}}
\newcommand{\cont}{\nonumber\eea\bea}
\newcommand{\cl}[1]{\begin{center} {#1} \end{center}}
\newcommand{\ba}{\begin{array}}
\newcommand{\ea}{\end{array}}

\newcommand{\ab}{{\alpha\beta}}
\newcommand{\cd}{{\gamma\delta}}
\newcommand{\dc}{{\delta\gamma}}
\newcommand{\ac}{{\alpha\gamma}}
\newcommand{\bd}{{\beta\delta}}
\newcommand{\abc}{{\alpha\beta\gamma}}
\newcommand{\eps}{{\epsilon}}
\newcommand{\lam}{{\lambda}}
\newcommand{\mn}{{\mu\nu}}
\newcommand{\mpnp}{{\mu'\nu'}}
\newcommand{\Amuu}{{A_{\mu}}}
\newcommand{\Amuo}{{A^{\mu}}}
\newcommand{\Vmuu}{{V_{\mu}}}
\newcommand{\Vmuo}{{V^{\mu}}}
\newcommand{\Anuu}{{A_{\nu}}}
\newcommand{\Anuo}{{A^{\nu}}}
\newcommand{\Vnuu}{{V_{\nu}}}
\newcommand{\Vnuo}{{V^{\nu}}}
\newcommand{\Fmnu}{{F_{\mu\nu}}}
\newcommand{\Fmno}{{F^{\mu\nu}}}

\newcommand{\abcd}{{\alpha\beta\gamma\delta}}


\newcommand{\bsigma}{\mbox{\boldmath $\sigma$}}
\newcommand{\btau}{\mbox{\boldmath $\tau$}}
\newcommand{\brho}{\mbox{\boldmath $\rho$}}
\newcommand{\bpipi}{\mbox{\boldmath $\pi\pi$}}
\newcommand{\bss}{\bsigma\!\cdot\!\bsigma}
\newcommand{\btt}{\btau\!\cdot\!\btau}
\newcommand{\bnabla}{\mbox{\boldmath $\nabla$}}
\newcommand{\bphi}{\mbox{\boldmath $\tau$}}
\newcommand{\bvarphi}{\mbox{\boldmath $\rho$}}
\newcommand{\bDelta}{\mbox{\boldmath $\Delta$}}
\newcommand{\bpsi}{\mbox{\boldmath $\psi$}}
\newcommand{\bPsi}{\mbox{\boldmath $\Psi$}}
\newcommand{\bPhi}{\mbox{\boldmath $\Phi$}}
\newcommand{\bnab}{\mbox{\boldmath $\nabla$}}
\newcommand{\bpi}{\mbox{\boldmath $\pi$}}
\newcommand{\btheta}{\mbox{\boldmath $\theta$}}
\newcommand{\bkappa}{\mbox{\boldmath $\kappa$}}

\newcommand{\bA}{{\bf A}}
\newcommand{\bB}{\mbox{\boldmath $B$}}
\newcommand{\bp}{\mbox{\boldmath $p$}}
\newcommand{\bk}{\mbox{\boldmath $k$}}
\newcommand{\bq}{\mbox{\boldmath $q$}}
\newcommand{\bfe}{{\bf e}}
\newcommand{\bb}{\mbox{\boldmath $b$}}
\newcommand{\br}{\mbox{\boldmath $r$}}
\newcommand{\bR}{\mbox{\boldmath $R$}}
\newcommand{\bs}{\mbox{\boldmath $s$}}

\newcommand{\fph}{${\cal F}$}
\newcommand{\aph}{${\cal A}$}
\newcommand{\dph}{${\cal D}$}
\newcommand{\fpi}{f_\pi}
\newcommand{\mpi}{m_\pi}
\newcommand{\Tr}{{\mbox{\rm Tr}}}
\def\Qb{\overline{Q}}
\newcommand{\delu}{\partial_{\mu}}
\newcommand{\delo}{\partial^{\mu}}
%
%
\newcommand{\up}{\!\uparrow}
\newcommand{\upup}{\uparrow\uparrow}
\newcommand{\updo}{\uparrow\downarrow}
\newcommand{\uu}{$\uparrow\uparrow$}
\newcommand{\ud}{$\uparrow\downarrow$}
\newcommand{\auu}{$a^{\uparrow\uparrow}$}
\newcommand{\aud}{$a^{\uparrow\downarrow}$}
\newcommand{\pu}{p\!\uparrow}

\newcommand{\qp}{quasiparticle}
\newcommand{\sa}{scattering amplitude}
\newcommand{\ph}{particle-hole}
\newcommand{\qcd}{{\it QCD}}
\newcommand{\integ}{\int\!d}
\newcommand{\ie}{{\sl i.e.~}}
\newcommand{\etal}{{\sl et al.~}}
\newcommand{\etc}{{\sl etc.~}}
\newcommand{\rhs}{{\sl rhs~}}
\newcommand{\lhs}{{\sl lhs~}}
\newcommand{\eg}{{\sl e.g.~}}
\newcommand{\ef}{\epsilon_F}
\newcommand{\sigt}{d^2\sigma/d\Omega dE}
\newcommand{\sige}{{d^2\sigma\over d\Omega dE}}
\newcommand{\rpaeq}{\beq
\left ( \begin{array}{cc}
A&B\\
-B^*&-A^*\end{array}\right )
\left ( \begin{array}{c}
X^{(\kappa})\\Y^{(\kappa)}\end{array}\right )=E_\kappa
\left ( \begin{array}{c}
X^{(\kappa})\\Y^{(\kappa)}\end{array}\right )
\eeq}
\newcommand{\ket}[1]{| {#1} \rangle}
\newcommand{\bra}[1]{\langle {#1} |}
\newcommand{\ave}[1]{\langle {#1} \rangle}
\newcommand{\half}{{1\over 2}}

\newcommand{\singlespace}{
    \renewcommand{\baselinestretch}{1}\large\normalsize}
\newcommand{\doublespace}{
    \renewcommand{\baselinestretch}{1.6}\large\normalsize}
\newcommand{\bftau}{\mbox{\boldmath $\tau$}}
\newcommand{\bfalpha}{\mbox{\boldmath $\alpha$}}
\newcommand{\bfgamma}{\mbox{\boldmath $\gamma$}}
\newcommand{\bfxi}{\mbox{\boldmath $\xi$}}
\newcommand{\bfbeta}{\mbox{\boldmath $\beta$}}
\newcommand{\bfeta}{\mbox{\boldmath $\eta$}}
\newcommand{\bfpi}{\mbox{\boldmath $\pi$}}
\newcommand{\bfphi}{\mbox{\boldmath $\phi$}}
\newcommand{\bfR}{\mbox{\boldmath ${\cal R}$}}
\newcommand{\bfL}{\mbox{\boldmath ${\cal L}$}}
\newcommand{\bfM}{\mbox{\boldmath ${\cal M}$}}
\def\dblint{\mathop{\rlap{\hbox{$\displaystyle\!\int\!\!\!\!\!\int$}}
    \hbox{$\bigcirc$}}}
\def\ut#1{$\underline{\smash{\vphantom{y}\hbox{#1}}}$}

\def\UNITY{{\bf 1\! |}}
\def\Pom{{\bf I\!P}}
\def\lsim{\mathrel{\rlap{\lower4pt\hbox{\hskip1pt$\sim$}}
    \raise1pt\hbox{$<$}}}         
\def\gsim{\mathrel{\rlap{\lower4pt\hbox{\hskip1pt$\sim$}}
    \raise1pt\hbox{$>$}}}         
\def\beq{\begin{equation}}
\def\eeq{\end{equation}}
\def\bea{\begin{eqnarray}}
\def\eea{\end{eqnarray}}


\title{Nonlinear $k_{\perp}$-factorization for
 Gluon-Gluon Dijets Produced off Nuclear Targets}%

\author{N.N. Nikolaev}%
\email{N. Nikolaev@fz-juelich.de}
\affiliation{Institut f\"ur Kernphysik, Forschungszentrum J\"ulich, D-52425 J\"ulich, Germany}
\affiliation{L.D. Landau Institute for Theoretical Physics, Moscow 117940, Russia}
\author{W. Sch\"afer}%
\email{Wo.Schaefer@fz-juelich.de}
\affiliation{Institut f\"ur Kernphysik, Forschungszentrum J\"ulich, D-52425 J\"ulich, Germany}
\author{B.G. Zakharov}%
\email{B.Zakharov@fz-juelich.de} \affiliation{L.D. Landau
Institute for Theoretical Physics, Moscow 117940, Russia}
\affiliation{Institut f\"ur Kernphysik, Forschungszentrum
J\"ulich, D-52425 J\"ulich, Germany}

\date{\today}%

\begin{abstract}
The origin of the breaking of conventional linear $k_{\perp}$-factorization
for hard processes in a nuclear environment is by now well established. The
realization of the nonlinear nuclear $k_{\perp}$-factorization which emerges instead
was found to change from one jet observable to another. Here we report on an
important technical progress, the evaluation of the four-gluon color dipole
cross section operator. It describes the  coupled seven-channel non-Abelian
intranuclear evolution of the four-gluon color-singlet states. An exact
diagonalization of this seven-channel problem is possible for large
number of colors $N_c$ and allows a formulation of nonlinear $k_\perp$-factorization
for production of gluon-gluon dijets. The momentum spectra for
dijets in all possible color representations are reported in the form 
of explicit quadratures in terms of the collective nuclear unintegrated glue.
Our results fully corroborate the concept of universality classes. 
\end{abstract}
\maketitle


\section{Introduction}

The key point behind conventional perturbative Quantum Chromo Dynamics
(pQCD) factorization theorems is that parton densities are low and a
single parton from the beam and single parton from the target 
participate in a hard reaction. As a result, hard cross 
sections are linear functionals (convolutions) of the appropriate parton 
densities in the projectile and target \cite{Textbook}. For instance, once
the unintegrated gluon density of the target proton is determined
from the deep inelastic scattering (DIS) structure function, it 
would allow a consistent description of all
other small-$x$, i.e., high-energy, processes of hard production 
off free nucleons. In contrast to that,
in hard production off nuclei the contributions of multigluon exchanges 
with the nucleus are enhanced by a large size of the target.
The principal consequence is a dramatic breaking of the conventional 
linear $k_\perp$-factorization for hard processes in a nuclear
environment which, according to the recent extensive studies 
\cite{Nonlinear,PionDijet,SingleJet,Nonuniversality,QuarkGluonDijet},
must be replaced by a nonlinear $k_\perp$-factorization.
Namely, one can take diffractive dijet production \cite{NZsplit,NSSdijet}
as a reference process for the definition of the collective 
nuclear unintegrated gluon density. Then, it turns out that
the so-defined nuclear glue furnishes the  familiar linear 
$k_{\perp}$-factorization description of the nuclear structure function
$F_{2A}(x,Q^2)$ and of the forward single-quark spectrum in DIS
(although the  linear $k_{\perp}$-factorization property of both
observables is
rather an exception due to the Abelian feature of the photon).
Furthermore, the dijet spectra in DIS and single-jet spectra in 
hadron-nucleus collisions admit a description in terms of the 
same collective nuclear gluon density, albeit in the form 
of highly nonlinear quadratures. The universality classes introduced
in \cite{Nonuniversality,QuarkGluonDijet} allow to relate the
nonlinearity properties of final states from different partonic
pQCD subprocesses to the pattern of color flow from the incident
parton to final-state dijet.  A full derivation of nonlinear
$k_\perp$-factorization for all high-energy single-jet spectra 
was published in \cite{SingleJet}; the forward quark-antiquark 
dijet production in DIS and pion-nucleus collisions was studied 
in \cite{Nonlinear} and \cite{PionDijet}, respectively; the
results for the two--particle pectrum of 
open heavy flavor  production $g\to Q\overline{Q}$ 
in gluon-nucleus collisions -- the dominant source of charm in proton-nucleus
collisions -- were presented in \cite{Nonuniversality,QuarkGluonDijet};
quark-gluon dijets in quark-nucleus interactions  -- the
dominant source of forward dijets in the proton fragmentation 
region of proton-nucleus collisions -- were treated in 
\cite{QuarkGluonDijet}. 

In this communication we report the derivation of nonlinear 
$k_\perp$-factorization for the last missing pQCD subprocess -
the production of hard gluon-gluon dijets in gluon-nucleus collisions
when the nuclear coherency condition $x\lsim x_A \approx 
0.1\cdot A^{-1/3}$ holds (for the definition
of $x_A$ for a target nucleus of mass number $A$ see below).
At the not so high  energies of the Relativistic Heavy Ion Collider (RHIC),  
the coherency condition can only be met in the proton fragmentation
region of $pA$ collisions, where the contribution from gluon-gluon 
dijets is  marginal (\cite{Kretzer} and references therein).
This subprocess
will be a principal building block of the pQCD description of
mid-rapidity dijet production in $pA$ collisions at the Large 
Hadron Collider (LHC), however. The non-Abelian intranuclear evolution 
of gluon-gluon dijets is quite involved - at arbitrary number
of colors $N_c$ two gluons couple to seven irreducible representations.
Based on the reduction of the dijet production problem to interaction 
with the nuclear target of color-singlet multiparton states 
\cite{SlavaPositronium,NPZcharm,SlavaLPM,Nonlinear,QuarkGluonDijet},
we report an explicit form of evolution matrices for arbitrary $N_c$.
We demonstrate how the forbidding case of seven-channel
non-Abelian evolution equations can be diagonalized in an explicit
form in the large-$N_c$ approximation - this is reminiscent of our
finding of the
reduction of the three-channel non-Abelian evolution for quark-gluon
dijets to a two-channel problem \cite{QuarkGluonDijet}. 

The production of gluon-gluon dijets on nuclear targets in a limit
of strong ordering of the rapdities of the produced gluons has 
been discussed earlier by several authors \cite{Kovchegov,Baier}. 
The starting points are similar, but these works stopped short 
of the explicit diagonalization of their version of the non-Abelian evolution
for the four-parton state. Baier et al. reported some numerical
results for the equal-momentum dijets. As we commented in
\cite{QuarkGluonDijet}, our analytical
results anticipated the trends of nuclear decorrelation
of dijets reported in \cite{Baier}. In contrast to \cite{Kovchegov,Baier},
our results for the  gluon-gluon dijets in all color representations 
are presented in the form of explicit quadratures in terms
of the collective nuclear unintegrated glue and do not assume 
a strong ordering of the gluon rapidities.
They fully confirm our concept of universality classes
\cite{Nonuniversality,QuarkGluonDijet}. For instance, the 
nonlinear $k_\perp$ factorization properties of excitation of 
digluons in higher color representations are the same as  
those in excitation of color-octet quark-antiquark dijets in
DIS and quark-gluon dijets in higher color
representations in $qA$ collisions. The only difference is in the collective
nuclear glue - different pQCD subprocesses pick up different
components of the color-density matrix for nuclear glue. 
This is also the case for excitation of dijets in the same color 
representation as the incident parton: $g\to \{gg\}_8,\, g\to \{q\bar{q}\}_8,
\, q\to \{qg\}_3$. The
diffractive excitation of digluons in the antisymmetric
octet is similar to diffractive excitation of  color-triplet $qg$ 
dijets in $qA$ collisions and color-octet quark-antiquark dijets
in $gA$ collisions. In both $g\to gg$ and $q\to qg$ processes 
coherent diffractive excitation of incident partons
with net color charge is suppressed by a nuclear absorption 
factor which can be identified with Bjorken's gap survival 
probability \cite{Bjorken}.  

The further presentation is organized as follows. We start
with the discussion of the reaction kinematics and the
master formula for the dijet cross section in Sec. 2. The
interaction properties of the two-gluon and three-gluon states
are presented in Sec. 3. The technically rather involved
derivation of the nuclear $\textsf{S}$-matrix for the four-gluon
state is the subject of Sec. 4. In Sec. 5 we report the 
linear $k_\perp$-factorization formula for the gluon-gluon 
dijet cross section for the free-nucleon target. The
principal new results of our study  -- nonlinear
 $k_\perp$-factorization formulas for gluon-gluon dijets
in different color representations, their classification
in universality classes and a comparison to other dijet
processes -- are reported in Sec. 6. In Conclusions we
summarize our main results.

The technicalities of the construction of the
irreducible representations
for the two-gluon states at an arbitrary number of colors $N_c$
are reported in Appendices A,B. 
The exact integration of non-Abelian evolution equations 
for the four-gluon system and the derivation of explicit quadratures 
for the dijet spectrum is possible only for large $N_c$,
although the calculation of higher order terms of $1/N_c$
perturbation theory is not a problem \cite{Nonlinear}. On
the other hand, the single-jet problem can be solved exactly 
at arbitrary $N_c$ \cite{SingleJet}, and in Appendix C we show
how the coupled seven-channel equations can be exactly
diagonalized in the $t$-channel basis appropriate for the
single-jet problem.  In Appendix D 
we give a summary of different 
components of the color-density matrix for nuclear glue
which enter the description of different pQCD subprocesses.


\section{The master formula for gluon-gluon dijet production off free nucleons and nuclei}

\subsection{Kinematics and nuclear coherency}

Our exposition of the master formula for dijet production follows
closely our recent work on quark--gluon dijets \cite{QuarkGluonDijet}.

To the lowest order in pQCD the underlying subprocess for gluon-gluon
dijet production in the proton fragmentation region of proton-nucleus
collisions is a collision of a gluon $g^*$ from the proton with a
gluon $g_N$ from the target,
$$
g^* g_N \to g g\,.
$$
It is a pQCD Bremsstrahlung off a gluon tagged by the scattered gluon.
We do not restrict ourselves to the emission of slow, $z \ll 1$ gluons.
In the case of a nuclear target one has to deal with
multiple gluon exchanges which are enhanced by a large thickness of the
target nucleus.

\begin{figure}[!t]
\begin{center}
\includegraphics[width = 5.5cm, height= 6.0cm]{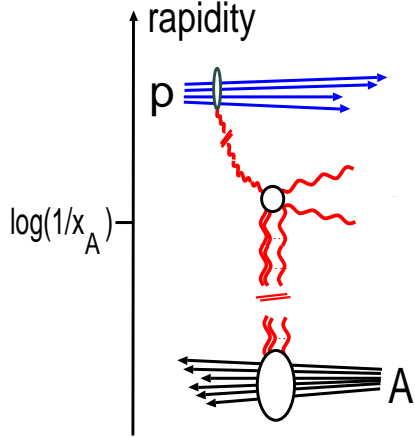}
\caption{The rapidity structure of the
radiation of gluons by gluons $g\to gg$ in the nuclear coherency region of $pA$
collisions.}
\label{fig:RapidityGluonGluon}
\end{center}
\end{figure}

From the laboratory, i.e., the nucleus rest frame, standpoint it can
be viewed as an excitation of the perturbative $|gg\rangle$ Fock state
of the physical projectile $|g^*\rangle$ by one-gluon exchange with the
target nucleon or multiple gluon exchanges with the target nucleus.
Here the collective nuclear effects develop, and the frozen impact
parameter approximation holds, if the coherency over
the thickness of the nucleus holds for the $gg$ Fock states, i.e.,
if the coherence length is larger than the diameter of the nucleus $2R_A$,
\beq
l_c ={ 2E_{g^*} \over (Q^*)^2 + M_{\perp}^2} = {1\over x m_N}> 2R_A\,,
\label{eq:2.1.1}
\eeq
where
\beq
M_{\perp}^2 = {\bp_1^2 \over z_1}+{\bp_2^2\over z_2}
\label{eq:2.1.2}
\eeq
is the transverse mass squared of the $gg$ state,
$\bp_{1,2}$ and $z_{1,2}$ are the transverse momenta and
fractions of the the incident gluon's momentum carried by the
outgoing gluon one and gluon two, respectively ($z_1+z_2=1$). 
The virtuality of the gluon $g^*$ equals $(Q^*)^2=(\bp^*)^2$, where $\bp^*$ is the
transverse momentum of $g^*$ in the incident proton
(Fig.~\ref{fig:RapidityGluonGluon}).
In the antilaboratory (Breit) frame, partons with a momentum
$xp_N$ have a longitudinal localization of the order of their
Compton wavelength $\lambda = 1/xp_N$, where $p_N$ is the momentum
per nucleon. The coherency over the thickness of the nucleus
in the target rest frame is equivalent to the spatial overlap of
parton fields of different nucleons at the same impact parameter
in the Lorentz-contracted ultrarelativistic nucleus. In the
overlap regime one would think of the fusion of partons
form different nucleons and collective nuclear
parton densities \cite{NZfusion}. The overlap takes place
if $\lambda$ exceeds the Lorentz-contracted thickness
of the ultrarelativistic nucleus,
\beq
\lambda = {1\over xp_N} > 2R_A \cdot {m_N\over p_N}\, ,
\label{eq:2.1.3}
\eeq
which is identical to the condition (\ref{eq:2.1.1}).

Qualitatively, both descriptions of collective nuclear effects
are equivalent to each other. Quantitatively, the
laboratory frame approach takes advantage of the
well developed multiple-scattering theory of
interactions of color dipoles with nuclei
\cite{NZ91,NZ92,NZ94,Nonlinear}.
From the practical
point of view, the coherency condition
$ x< x_A $ restricts collective effects in hard processes at RHIC
to the proton fragmentation region of $pA,dA$ collisions, but at LHC 
our treatment will hold down to the mid-rapidity region of $pA$
collisions.
The target frame rapidity structure of the considered $g^*\to gg$
excitation
is shown in Fig.~\ref{fig:RapidityGluonGluon}. The (pseudo)rapidities
of the final state partons must satisfy  $\eta_{1,2} > \eta_A=
\log{1/x_A}$. The rapidity separation of the two hard gluon jets,
\beq
\Delta\eta_{gg} = \log {z_2\over z_1}\,,
\label{eq:2.1.4}
\eeq
is considered to be finite. Both jets are supposed to be
separated by a large rapidity from other jets at mid-rapidity
or in the target nucleus hemisphere; the gaps between all jets,
beam spectators and target debris are filled by soft hadrons
from an underlying event. 

Clearly, the incident gluon $g^*$ is a parton of the color-singlet 
beam hadron and is accompanied by comoving spectators. However, the 
effect of interactions of comoving spectator
partons cancels out upon the fully inclusive integration over
the spectator phase space \cite{NPZcharm}, an explicit
demonstration of such a cancellation is found in \cite{SingleJet}. 
The properties of the beam hadrons only define the longitudinal and transverse
momentum spectrum of $g^*$ in the beam, and the problem we treat
here will be a building block of the pQCD description of mid-rapidity
to proton hemisphere
dijets in $pA$ collisions at LHC. 
 

\subsection{Master formula for excitation of gluon-gluon dijets}

In the nucleus rest frame, relativistic partons
$g^*,g_1$ and $g_2$,
propagate along straight-line, fixed-impact-parameter, trajectories.
To the lowest
order in pQCD the Fock state expansion for
the physical state $|g^*\rangle_{phys}$ reads
\beq
 \ket{g^*}_{phys} = \ket{g^*}_0 + \Psi(z_1,\br) \ket{gg}_0\, ,
\label{eq:2.2.1}
\eeq
where $\Psi(z_1,\br)$ is the probability amplitude to find the $gg$ system
with the separation $\br$ in the two-dimensional impact parameter space,
the subscript $"0"$ refers to bare partons.
The perturbative
coupling of the $g^*\to gg$ transition is reabsorbed into the lightcone
wave function $\Psi(z_1,\br)$.
We also omitted a wave function renormalization factor,
which is of no relevance for the inelastic excitation
to the perturbative order discussed here. The explicit
expression for  $\Psi(z_1,\br)$ in terms of the gluon-splitting
function and the gluon virtuality  $(Q^*)^2$ 
will be presented below. For the
sake of simplicity we take the collision axis along the momentum of
the incident quark $g^*$, the transformation between the transverse momenta in
the $g^*$-target and $p$-target reference frames is trivial
\cite{SingleJet}.

\begin{figure}[!t]
\begin{center}
\includegraphics[width = 5.5cm,height=13cm, angle = 270]{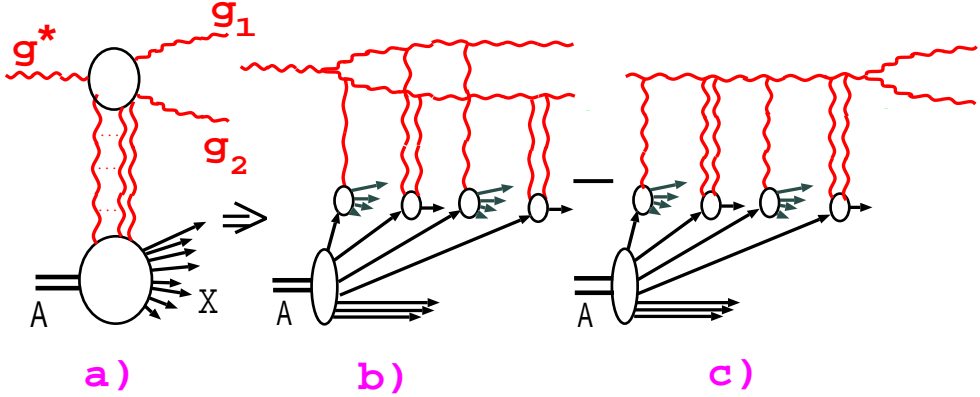}
\caption{Typical contribution to the excitation amplitude for $g A \to g_1 g_2 X$,
with multiple color excitations of the nucleus.
The amplitude receives contributions from processes that involve interactions
with the nucleus after and before the virtual
decay which interfere destructively.}
\label{fig:GGDijetDiagram}
\end{center}
\end{figure}

If $\bb$ is the impact parameter of the projectile $g^*$, then
\beq
\bb_{1}=\bb-z_2\br, \quad\quad \bb_{2}=\bb+z_1\br\, .
\label{eq:2.2.2}
\eeq
By the conservation of
impact parameters, the action of the $\textsf{S}$-matrix on $\ket{a}_{phys}$
takes a simple form
\bea
&& \textsf{S}\ket{g^*}_{phys} =
\textsf{S}_g(\bb) \ket{g^*}_0 +
\textsf{S}_g(\bb_1) \textsf{S}_g(\bb_2)\Psi(z_1,\br) \ket{gg}_0 \nonumber \\
&&=\textsf{S}_g(\bb) \ket{g^*}_{phys} +  
[ \textsf{S}_g(\bb_1) \textsf{S}_g(\bb_2) - \textsf{S}_g(\bb) ]
\Psi(z_1,\br) \ket{gg}_0 \, .
\label{eq:2.2.3} 
\eea
In the last line we
explicitly decomposed the final state into the (quasi)elastically
scattered $\ket{g^*}_{phys}$ and the excited state $\ket{gg}_{0}$.
The two terms in the latter describe a scattering on the target of
the $gg$ system formed way in front of the target and the
transition $g^*\to gg$ after the interaction of the state
$\ket{g^*}_0$ with the target, as illustrated in Fig.
\ref{fig:GGDijetDiagram}.
The contribution from transitions
$g^*\to gg$ inside the target nucleus vanishes in the high-energy
limit of $x \lsim x_A$ \footnote{In terms of the lightcone approach
to the QCD Landau-Pomeranchuk-Migdal effect, this corresponds to
the thin-target limit \cite{SlavaLPM}.}. We recall, that the $s$-channel helicity of all
gluons is conserved.

\begin{figure}[!t]
\begin{center}
\includegraphics[width = 5.5cm,angle=270]{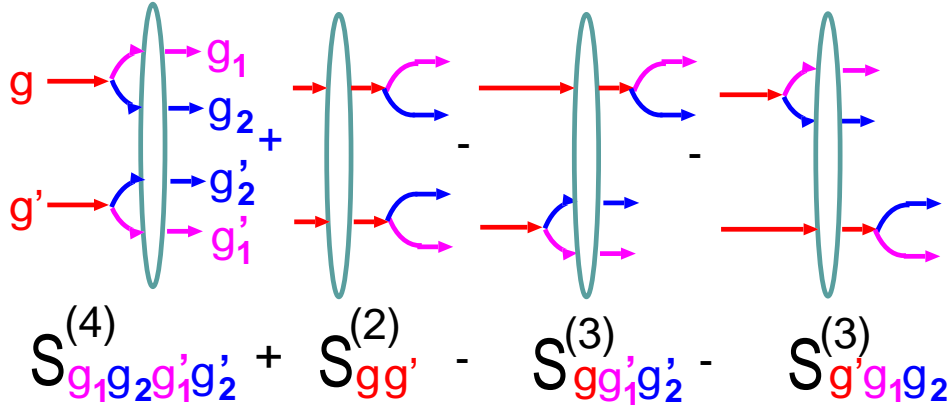}
\caption{The $\textsf{S}$-matrix structure of the two-body density
matrix for excitation $g\to g_1g_2$.}
\label{fig:Distortions3gluon}
\end{center}
\end{figure}

The probability amplitude for the two-jet spectrum is given by the
Fourier transform
\beq \int d^2\bb_1 d^2\bb_2 \exp[-i(\bp_1\bb_1 +
\bp_2\bb_2)][ \textsf{S}_g(\bb_1) \textsf{S}_g(\bb_2) - \textsf{S}_g(\bb) ] \Psi(z_1,\br)
\label{eq:2.2.4} \eeq
The differential cross section is proportional
to the modulus squared of (\ref{eq:2.2.4}),
\bea
&&\int d^2\bb'_1 d^2\bb'_2 \exp[i(\bp_1\bb'_1 +
\bp_2\bb'_2)][ \textsf{S}_g^{\dagger}(\bb'_1) \textsf{S}_g^{\dagger}(\bb'_2) - 
\textsf{S}_g^{\dagger}(\bb') ] \Psi^{*}(z_1,\br')\nonumber\\
&\times& \int d^2\bb_1 d^2\bb_2 \exp[-i(\bp_1\bb_1 +
\bp_2\bb_2)][ \textsf{S}_g(\bb_1) \textsf{S}_g(\bb_2) - \textsf{S}_g(\bb) ] \Psi(z_1,\br)\, .
\label{eq:2.2.5} \eea
The crucial point is that
the hermitian conjugate
$\textsf{S}^{\dagger}$ can be viewed as the $\textsf{S}$-matrix for an antiparton
\cite{SlavaPositronium,NPZcharm,Nonlinear}. Consequently,
the four terms in the product
$$[ \textsf{S}_g(\bb'_1) \textsf{S}_g(\bb'_2) - \textsf{S}_{g}(\bb') ]^\dagger
[ \textsf{S}_g (\bb_1) \textsf{S}_g(\bb_2) - \textsf{S}_g(\bb) ]
$$
admit a simple interpretation:
\bea
\textsf{S}^{(2)}_{{g^*}' g^*}(\bb',\bb)&=& \textsf{S}_{g}^{\dagger}(\bb')\textsf{S}_g(\bb)
\label{eq:2.2.6}
\eea
can be viewed as an $\textsf{S}$-matrix for elastic scattering on a
target of the ${g^*}'g^*$ state in which the (anti)gluon
${g^*}'$ propagates at the impact parameter $\bb'$. The averaging
over the color states of the beam parton $g^*$ amounts to the
dipole $g^* g^*$ being in the color singlet state. 
Similarly,
\bea
 \textsf{S}^{(3)}_{{g^*}'g_1g_2}(\bb',\bb_1,\bb_2) &=&
\textsf{S}_g^{\dagger}(\bb')\textsf{S}_g(\bb_1) \textsf{S}_g(\bb_2), \nonumber \\
\textsf{S}^{(3)}_{g'_1 g'_2 g^*}(\bb,\bb'_1,\bb'_2) &=&
\textsf{S}_g^{\dagger}(\bb'_1)\textsf{S}_g^{\dagger}(\bb'_2) \textsf{S}_g(\bb)
 \nonumber \\
\textsf{S}^{(4)}_{g'_1 g'_2  g_1  g_2}(\bb'_1,\bb'_2,\bb_1,\bb_2) &=&
\textsf{S}_g^{\dagger}(\bb'_1)\textsf{S}_g^{\dagger}(\bb'_2) \textsf{S}_g(\bb_1)\textsf{S}_g(\bb_2) \,
.
\label{eq:2.2.7}
\eea
describe elastic scattering on a
target of the overall
color-singlet three- and four-gluon states,
respectively. This is shown schematically in
Fig.~\ref{fig:Distortions3gluon}.
Here we suppressed the matrix elements of $\textsf{S}^{(n)}$ over the
target nucleon, for details of the derivation based on the
closure relation, see \cite{Nonlinear}. Specifically, in the
calculation of the inclusive cross sections one averages over
the color states of the beam gluon $g$, sums over color
states $X$ of final state gluons $q_1,g_2$, takes the matrix products of
$\textsf{S}^{\dagger}$ and $\textsf{S}$ with respect to the relevant color indices
entering $\textsf{S}^{(n)}$ and sums over all nuclear
final states applying the closure relation.
The technicalities of the derivation
of $\textsf{S}^{(n)}$ will be presented below, here we cite
the master formula for the dijet cross section, which is the Fourier
transform of the two-body density matrix:
\bea
&&{d \sigma (g^* \to g_1g_2) \over dz d^2\bp_1 d^2\bp_2 } = {1 \over (2 \pi)^4} \int
d^2\bb_1 d^2\bb_2 d^2\bb'_1
 d^2\bb'_2 \nonumber \\
&&\times \exp[-i \bp_2
(\bb_2 -\bb'_2) - i
\bp_1(\bb_1
-\bb'_1) ] \Psi(z_1,\bb_1 -
\bb_2) \Psi^*(z_1,\bb'_1-
\bb'_2) \\
&&\sum_{X}\bra{X}\Biggl\{
\textsf{S}^{(4)}_{g'_1g'_2 g_1 g_2}(\bb'_1,\bb'_2,\bb_1,\bb_2)
+ \textsf{S}^{(2)}_{{g^*}'g^*}(\bb',\bb) -
\textsf{S}^{(3)}_{g'_1 g'_2 g^*}(\bb,\bb'_1,\bb'_2)
- \textsf{S}^{(3)}_{{g^*}'g_1 g_2}(\bb',\bb_1,\bb_2) \Biggr\}\ket{in}\nonumber
\label{eq:2.2.8}
\eea

Hereafter, we describe the final state dijet in terms of the jet momentum
 $\bp\equiv \bp_1,~~z\equiv z_1$,
and the decorrelation (acoplanarity) momentum
$\bDelta=\bp_1+\bp_2$. We also introduce
\beq
\bs=\bb_2-\bb'_2\,,
\label{eq:2.2.9}
\eeq
in terms of which $\bb_1-\bb'_1=\bs+\br -\br'$ and
\bea
 \exp[-i \bp_2
(\bb_2 -\bb'_2) - i
\bp_1(\bb_1
-\bb'_1)] =\exp[-i\bDelta \bs- i\bp \br +i\bp \br']\, ,
\label{eq:2.2.10}
\eea
so that the dipole parameter $\bs$ is conjugate to the acoplanarity
 momentum $\bDelta$.

\section{Calculation of the 2-parton and 3-parton $\textsf{S}$-matrices}

\subsection{The gluon-nucleon $\textsf{S}$-matrix and the $k_{\perp}$-factorization
representations for the color dipole cross section}\

\begin{figure}[!t]
\begin{center}
\includegraphics[width = 12.5cm]{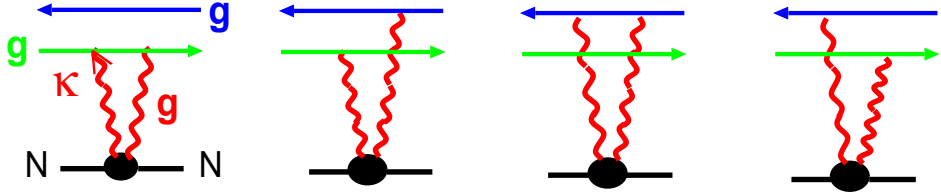}
\caption{The four Feynman diagrams for the gluon-gluon
dipole-nucleon interaction by the two-gluon pomeron exchange
in the $t$-channel.}
\label{fig:GluonGluonSigma}
\end{center}
\end{figure}

In order to set up the formalism, we start with the $\textsf{S}$-matrix
representation for the cross section of interaction
of the $gg$ color dipole with the free-nucleon
target.
To the two-gluon exchange approximation, the $\textsf{S}$-matrix of the
gluon-nucleon interaction equals
\bea
\textsf{S}_N(\bb) & =& 1
+ iT^a V_a\chi(\bb)- {1\over 2}
T^aT^a \chi^2(\bb)\, , 
\label{eq:3.1.1}
\eea
where $T^a$ is the $SU(N_c)$ generator in the adjoint 
representation, $\bra{g^b} T^a \ket{g^c} = -i f_{abc}$,
and $T^a V_a\chi(\bb)$ is the gluon-nucleon eikonal for single gluon
exchange. The vertex $V_a$ for excitation
of the nucleon $g^a N \to N^*_a$ into a color octet state is so
normalized that after application of closure over
the final state excitations $N^*$ the vertex $g^a g^b
NN$ equals $\bra{ N} V_a^\dagger V_b \ket{ N}=\delta_{ab}$.
The second order term in (\ref{eq:3.1.1}) already uses this normalization.
The $\textsf{S}$-matrix of the
$gg$-nucleon interaction equals
\beq
\textsf{S}^{(2)}_{gg}(\bb_1,\bb_2)=
{\bra{N}{\rm Tr}[\textsf{S}_N(\bb_1) \textsf{S}^\dagger_N(\bb_2)]
\ket{N} \over \bra{N} {\rm Tr} \openone  \ket{N}}\,.
\label{eq:3.1.2}
\eeq
The corresponding profile function is
$\Gamma_2(\bb_1,\bb_2)= 1 - \textsf{S}^{(2)}_{gg}(\bb_1,\bb_2)$.
The dipole cross section for interaction of the color-singlet
$gg$ dipole $\br=\bb_1-\bb_2$  with the free nucleon
is obtained upon the integration over the overall 
impact parameter
\bea
\sigma(\br) = 2\int d^2\bb_1
\Gamma_2(\bb_1,\bb_1-\br) =
C_A \int
d^2\bb_1 \Big[ \chi(\bb_1)-\chi(\bb_1-\br) \Big]^2\, ,
\label{eq:3.1.3}
\eea
where $C_A=T^aT^a = N_c$ is the gluon Casimir operator.
Eq. \ref{eq:3.1.3} sums up the contributions from
the four Feynman diagrams of Fig. \ref{fig:GluonGluonSigma}
and is related to the gluon density in the target by
the $k_{\perp}$-factorization formula \cite{NZ94,NZglue}
\bea
\sigma(x,\br) &=& {C_A \over C_F} \int d^2\bkappa f (x,\bkappa)
[1-\exp(i\bkappa\br)]\,,
\label{eq:3.1.4}
\eea
where $C_F = {(N_c^2-1)/2N_c}$ is the quark Casimir. Recall that the 
unintegrated gluon density 
\beq
{\cal F}(x,\kappa^2) = {\partial G(x, \kappa^2) \over \partial \log \kappa^2}
 \label{eq:3.1.5}
\eeq
was defined with respect to the $q\bar{q}$ color
dipole probe, and is related to $f(x,\bkappa)$ through 
\bea
f (x,\bkappa)&=& {4\pi \alpha_S(r)\over N_c}\cdot {1\over \kappa^4} \cdot {\cal
F}(x,\kappa^2) \, .
\label{eq:3.1.6}
\eea
Hereafter, unless it may cause confusion, we
suppress the variable $x$ in the gluon densities and
dipole cross sections.
The energy dependence of the dipole
cross section is governed by the color-dipole leading Log${1\over x}$ evolution 
\cite{NZZBFKL,NZ94}, the related evolution for the
unintegrated gluon density is described by the familiar
momentum-space BFKL (Balitsky-Fadin-Kuraev-Lipatov)  equation \cite{BFKL}.

The $\textsf{S}$-matrix for coherent interaction of the color dipole
with the nuclear target is given by the Glauber-Gribov
formula \cite{Glauber,Gribov}
\bea
\textsf{S}[\bb,\sigma(\br)]&=&\exp[-{1\over 2}\sigma(\br) T(\bb)]\, ,
\label{eq:3.1.7}
\eea
where
\bea
T(\bb)= \D \int_{-\infty}^\infty dr_z \, n_A(\bb,r_z)
\label{eq:3.1.8}
\eea
is the optical thickness
of the nucleus. The nuclear density $n_A(\bb,r_z)$ is
normalized according to $\D \int d^3\vec{r} \, n_A(\bb,r_z) = \int
d^2\bb\,  T(\bb) = A$, where $A$ is the nuclear mass number.

In the specific case of $\textsf{S}^{(2)}_{g^* g^*}(\bb',\bb)$ the
color dipole equals
\beq
\br_{gg}= \bb-\bb'=\bs +z\br-z\br'\,
\label{eq:3.1.9}
\eeq
and  $\textsf{S}^{(2)}_{g^*g^*}(\bb',\bb)$ entering Eq.~(\ref{eq:2.2.8})
will be given by the Glauber-Gribov formula
\beq
\textsf{S}^{(2)}_{g^*g^*}(\bb',\bb)=
\textsf{S}[\bb,\sigma(\bs +z\br-z\br')]\,.
\label{eq:3.1.10}
\eeq

\subsection{The $\textsf{S}$-matrix for the color-singlet $ggg$ system}

\begin{figure}[!t]
\begin{center}
\includegraphics[width = 10.5cm]{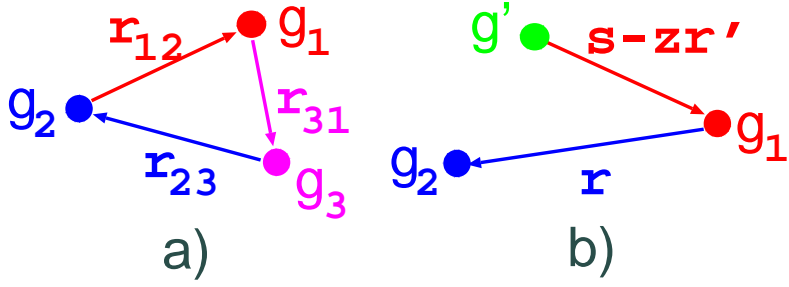}
\caption{The color dipole structure of (a) the generic 3-gluon
system of dipoles and (b) of the $g'g_1g_2$ system which emerges in the
 $\textsf{S}$-matrix structure of the two-body density
matrix for excitation $g\to g_1g_2$.}
\label{fig:QuarkAntiquarkGluonDipole}
\end{center}
\end{figure}

Here, there are two possibilities to couple three gluons to a
color singlet, but only the $f$-coupling is relevant to our problem,
see also Appendix C.

For the generic 3-gluon state
shown in Fig.~\ref{fig:QuarkAntiquarkGluonDipole} the color-dipole cross section 
equals
\bea
\sigma^{(3)}(\bb_1,\bb_2,\bb_3)= {1\over 2}\bigl[\sigma(\br_{12})+
\sigma(\br_{23})
+\sigma(\br_{31})\bigr],
\label{eq:3.2.1}
\eea
where $\br_{ik}=\bb_k-\bb_i$.
The configuration of color dipoles for the case of our interest
is shown in Fig.~\ref{fig:QuarkAntiquarkGluonDipole} (see the related derivation
in \cite{NPZcharm}). For the $g'q_1g_2$ state the relevant
dipole sizes in (\ref{eq:3.2.1}) equal
\bea
\br_{g_1{g^*}'}&=&\bb_1-\bb'=\bs-z\br\,,\nonumber\\
\br_{g_1g_2}&=&\bb_2-\bb_1=\br\,,\nonumber\\
\br_{{g^*}'g_2}&=&\bb'-\bb_2=\bs+\br-z\br',\
\label{eq:3.2.2}
\eea
so that
\bea
&&\sigma_{{g^*}'q_1g_2}={1\over 2}\bigl[\sigma(\br)+
\sigma(\bs+\br-z\br')
+\sigma(\bs-z\br')\bigr]\,, \nonumber\\
&&\sigma_{g^*g_1'g_2'}={1\over 2}\bigl[\sigma(-\br')+
\sigma(\bs-\br'+z\br)
+\sigma(\bs+z\br)\bigr]\,.
\label{eq:3.2.3}
\eea
The overall color-singlet 3-gluon state has a unique color structure
and  its elastic scattering on a nucleus
is a single-channel problem. Consequently, the nuclear $\textsf{S}$-matrix
is given by the single-channel Glauber-Gribov
formula 
\bea
\textsf{S}^{(3)}_{{g^*}'q_1g_2}(\bb,\bb_q',\bb_g')&=&\textsf{S}[\bb,\sigma_{{g^*}'q_1g_2}]\,,\nonumber\\
\textsf{S}^{(3)}_{g^*g_1'g_2'}(\bb',\bb_q,\bb_g)&=&\textsf{S}[\bb,\sigma_{g^*g_1'g_2'}]\,.
\label{eq:3.2.4}
\eea


\section{Dipole cross section operator for four-gluon states}

We now come to the major technical novelty of this paper, the dipole
cross section matrix for the four--gluon system.
Here, as in the previous applications for $qq\bar{q}\bar{q}$ and $q\bar{q}gg$
systems, the large $N_c$ limit offers a particularly useful expansion.
We thus discuss the general case in which the gluon is transforming
in the adjoint representation of $SU(N_c)$.
In order to construct  the relevant four-gluon states,
our first task is to decompose the product-states of the 
adjoint $\times$ adjoint system into irreducible representations.
The adjoint (or regular) representation of $SU(N_c)$ has
$N_c^2-1$ states, and the Clebsch-Gordan series for the product
of two adjoints reads
\bea
(N_c^2-1) \times (N_c^2-1)&=& 1 + (N_c^2-1)_A + (N_c^2-1)_S \nonumber  \\
&& + {(N_c^2-4)(N_c^2-1) \over 4} +
\Big[{(N_c^2-4)(N_c^2-1) \over 4}\Big]^* \nonumber \\
&& + {N_c^2 (N_c+3)(N_c-1) \over 4} + {N_c^2 (N_c-3)(N_c+1) \over 4} \nonumber \\
&=& 1 + 8_A + 8_S + 10 + \overline{10} + 27 + R_7 \, .
\label{eq:4.0.1}
\eea
In the last line, we named the representations by their $SU(3)$ dimensions,
except for one of the symmetric representations that vanishes for $N_c=3$,
and will be referred to as  $R_7$.


\subsection{Projection operators onto irreducible representations}

The derivation of projectors onto the representations (\ref{eq:4.0.1}) is
a lengthy, though standard exercise \cite{Predrag}.
While in our earlier solution
of the analogous problem for the $qgg\bar{q}$-system it had proven convenient
to represent gluons in a double line notation as pointlike
quark--antiquark systems, here we find it expedient to stick to
purely adjoint-index tensors. We present here a sketch of the construction
of irreducible representations (\ref{eq:4.0.1}), the details are given
in Appendices A and B

If $t^a, a = 1 \dots N_c^2-1$
are $SU(N_c)$-generators in the fundamental representation, 
the familiar $f$-- and $d$--tensors are defined through
\bea
t^a t^b = {1 \over 2 N_c} \delta_{ab} \, \openone 
+ { 1\over 2} \Big( d_{abc} + i f_{abc} \Big) t^c \, ,
\label{eq:4.1.1}
\eea
and
$ T^a_{bc} = -i f_{abc} $ are the $SU(N_c)$ generators in the adjoint representation.

First we decompose the product representation space into its 
symmetric and antisymmetric parts, respectively:
\bea
 \openone^{ab}_{cd} \equiv  \delta_{ac} \delta_{bd} = {\cal{S}}^{ab}_{cd} + {\cal{A}}^{ab}_{cd},
\label{eq:4.1.3}
\eea
where 
\bea
{\cal{S}}^{ab}_{cd}  \equiv {1 \over 2}\Big( \delta_{ac} \delta_{bd} + 
\delta_{ad} \delta_{bc} \Big)\,, \,\, 
{\cal{A}}^{ab}_{cd}  \equiv {1 \over 2}\Big( \delta_{ac} \delta_{bd} - 
\delta_{ad} \delta_{bc} \Big) \, .
\label{eq:4.1.2}
\eea
The complex
\bea
i Y^{ab}_{cd} && \equiv {i \over 2} \Big(  d_{adk}f_{kbc} +  f_{adk} d_{kbc} \Big)
\label{eq:4.1.4}
\eea
as well as 
\bea
[D_t]^{ab}_{cd} \equiv d_{ack}d_{kbd}\, , \, [D_u]^{ab}_{cd} \equiv d_{adk}d_{kbc} \, ,
[D_s]^{ab}_{cd} \equiv d_{abk}d_{kcd}
\label{eq:4.1.5}
\eea
also prove helpful. 
All the above defined tensors ${\cal{S}},{\cal{A}}, iY, D_s,D_t,D_u$
are hermitian.
        
The $SU(N_c)$-projectors into the singlet as well as the 
two adjoint multiplets have 
manifestly the same form as 
their well-known $N=3$ counterparts:
\bea
P[1]^{ab}_{cd} &=& {1 \over N_c^2-1} \delta_{ab} \delta_{cd} \\
P[8_A]^{ab}_{cd} &=& {1 \over N_c} f_{abk}f_{kcd} = {1 \over N_c} i f_{abk} i f_{kdc} \\
P[8_S]^{ab}_{cd} &=& {N_c \over N_c^2-4}  d_{abk}d_{kcd} =  {N_c \over N_c^2-4}\, [D_s]^{ab}_{cd}
\label{eq:4.1.7}
\eea
For the construction of higher multiplets a useful quantity
is
$
Q_{ab}^{cd} = 4 \cdot \Tr \Big[ t^a t^d t^b t^c \Big], 
$
which, with indices suppressed, equals
\bea
Q = { 1\over N_c} \Big( 2 {\cal{S}} - (N_c^2-1) P[1] \Big) + 
{1 \over 2} \Big( D_t + D_u - D_s \Big) + iY
\label{eq:4.1.8}
\eea
Then, the crucial observation is \cite{Predrag}, that
\bea
\openone - Q^2 = {N_c^2 -1 \over N_c} P[1] + {N_c^2 - 4 \over N_c} P[8_S] - P[8_A] \, .
\label{eq:4.1.9}
\eea
Apparently in the subspaces that are projected onto by
\bea
{\cal{S}}_\perp = {\cal{S}}- P[1] - P[8_S] \, , \,\, {\cal{A}}_\perp = {\cal{A}}- P[8_A] \, ,
\label{eq:4.1.10}
\eea
the operator $\openone - Q^2 = (\openone + Q)(\openone - Q)$ vanishes and
$Q$ has eigenvalues $\pm 1$.
We can thus write down
projection operators that decompose ${\cal{S}}_\perp,{\cal{A}}_\perp$
further, as
\bea
P_{A_\perp}^{\pm} = {1 \over 2} \, (\openone \pm Q )  {\cal{A}}_\perp ,\,\,
P_{S_\perp}^{\pm} = {1 \over 2} \, (\openone \pm Q )  {\cal{S}}_\perp\, .
\label{eq:4.1.11}
\eea
Checking how many states are contained in $P_{A_\perp}^{\pm},P_{S_\perp}^{\pm}$ ,
one may identify, that $P_{A_\perp}^{+} = P[10], P_{A_\perp}^{-} = P[\overline{10}], 
P_{S_\perp}^{+} = P[27],P_{A_\perp}^{-} =P[R_7] $. 
In convenient form, with all indices shown again, they read:
\bea
P[10]^{ab}_{cd}  &=& {1 \over 2} \Big( {\cal{A}}^{ab}_{cd} - P[8_A]^{ab}_{cd} + i Y^{ab}_{cd} \Big)
\label{eq:4.1.12}\nonumber\\
P[\overline{10}]^{ab}_{cd}  &=& {1 \over 2} \Big( {\cal{A}}^{ab}_{cd}
- P[8_A]^{ab}_{cd} - i Y^{ab}_{cd} \Big) \, .
\label{eq:4.1.13}
\eea
\bea
P[27]^{ab}_{cd} &=& {1 \over 2N_c} \Big( (N_c+2) {\cal{S}}^{ab}_{cd} - (N_c+2)(N_c-1) P[1]^{ab}_{cd}
\nonumber \\ &-& {1 \over 2} (N_c-2)(N_c+4) P[8_S]^{ab}_{cd}
 + {N_c\over 2} ([D_t]^{ab}_{cd}+ [D_u]^{ab}_{cd})
\Big) \label{eq:4.1.14}\\
P[R_7]^{ab}_{cd} & =& {1 \over 2N_c} \Big( (N_c-2) {\cal{S}}^{ab}_{cd} + (N_c-2)(N_c+1) P[1]^{ab}_{cd}
\nonumber \\ &+&  {1 \over 2} (N_c+2)(N_c-4) P[8_S]^{ab}_{cd}
 - {N_c\over 2} ([D_t]^{ab}_{cd}+ [D_u]^{ab}_{cd})\Big) \, .
\label{eq:4.1.15}
\eea

It is now a simple matter to obtain the quadratic Casimirs
(i.e. color charge squared) of the individual multiplets,
which can be found in Table I.

\begin{table}
\caption{Properties of Multiplets}
\begin{ruledtabular}
\begin{tabular}{lccccccc}
& \multicolumn{4}{|c|} {symmetric} &  \multicolumn{3}{c} {antisymmetric} \\
\hline
Name of rep.  &  $1$  &   $8_S$  &  $27$ &  $R_7$  &  $8_A$  &  $10$  &  $\overline{10}$ \\
 Dimension  &  1  & $N_c^2-1$  & ${N_c^2 (N_c+3)(N_c-1) \over 4}$ & 
${N_c^2 (N_c-3)(N_c+1) \over 4}$ & $ N_c^2-1$ & $ {(N_c^2-4)(N_c^2-1) \over 4} $ & 
$ {(N_c^2-4)(N_c^2-1) \over 4} $ \\
Casimir $C_2[R]$ & 0 & $N_c$ &  $2(N_c+1)$ &  $2(N_c-1)$ &  $N_c$ &  $2N_c$ &  $2N_c$ \\
$\lambda_R = 1- {C_2[R] \over 2C_A}$ & 1 & ${1 \over 2}$ & $-{1 \over N_c}$ & ${1 \over N_c}$ & 
${1 \over 2}$ & 0 & 0 \\      
\end{tabular}
\end{ruledtabular}
\label{table1}
\end{table}


\subsection{Multigluon states}

The above given projectors can be used to construct the color--space
wave function of the multigluon states relevant for us.
The four gluons have to be in a total color singlet, and
all possible states are exhausted by coupling a chosen pair of gluons
to all possible multiplets and the remaining two to an anti-multiplet.
The choice of pairs is of course arbitrary. Because averaging over 
colors of the incoming gluon amounts to the initial state
\bea \ket{\mathrm{in}} = {1 \over \sqrt{N_c^2-1}} \ket{8_A 8_A} 
 \, , 
\label{eq:4.2.1}\eea
a convenient choice is the $s$-channel pairing
\bea
\ket{R\overline{R}} && \equiv \ket{\Big\{ \Big[g^a(\bb_1) \otimes g^b(\bb_2) \Big]_R 
\otimes \Big[g^c(\bb'_1) \otimes g^d(\bb'_2) \Big]_{\overline{R}} \Big\}_1 } 
\nonumber \\ 
&&=  {1 \over \sqrt{\dim
[R]}} \,P[R]^{ab}_{cd} \, \ket{g^a(\bb_1) \otimes g^b(\bb_2) 
\otimes g^c(\bb'_1) \otimes g^d(\bb'_2)} \, .
\label{eq:4.2.2}\eea
The basis of color-singlet four-gluon states which contribute to the
non-Abelian evolution of gluon-gluon dijets in our problem consists of 
$\ket{11},\,\ket{8_A8_A},\,\ket{8_S8_S},\,\ket{10\overline{10}},\,
\ket{\overline{10}10},\,\ket{2727},\,\ket{R_7 R_7}.$ The mixed-symmetry color-singlet 
states like $\ket{8_A8_S}$ are possible but decouple from the above states.

We note in passing, that the single-jet spectrum derives from the dijet 
spectrum by integration over all $\bDelta$, which entails $\bs=0$. 
For studying the transition from the dijet to single-jet problem
an alternative, $t$-channel, pairing of gluons, $(g_1g_1')$ and $(g_2g_2')$,  
proves to be a more convenient one. Evidently,
the multiparton $\textsf{S}$-matrices for different choices are related
by a trivial permutation of the gluon impact parameters. For the reference 
purposes, in Appendix C we demonstrate how the coupled seven-channel 
problem is exactly diagonalized for the color-dipole configuration
appropriate to the single-jet problem.


\subsection{Multiparton S-matrix and the four-body color dipole cross section operator
for the free-nucleon target}

The frozen impact parameter approximation leads to a four-gluon S-matrix of the form
\bea
\textsf{S}^{(4)}_N(\bb_1,\bb_2,\bb'_1,\bb'_2) = \textsf{S}_N(\bb_1) \otimes \textsf{S}_N(\bb_2) 
\otimes \textsf{S}_N(\bb'_1) \otimes \textsf{S}_N(\bb'_2)\, .
\label{eq:4.3.1}\eea
Then, upon using eq.~(\ref{eq:3.1.1}), on a color-singlet four-particle state, 
the S-matrix $\textsf{S}_N^{(4)}$ takes the form
\bea
\textsf{S}_N^{(4)}(\bb_1,\bb_2,\bb'_1,\bb'_2) &=&  1 - 
{1 \over 2} \big[T^a_1 \chi(\bb_1) + T^a_2 \chi(\bb_2) 
+ T^a_{1'} \chi(\bb'_1) +  T^a_{2'} \chi(\bb'_2)\big]^2 \nonumber \\
&=& 1 - {1 \over 2} C_A \big[ \chi^2(\bb_1) +\chi^2(\bb_2)+\chi^2(\bb'_1)+\chi^2(\bb'_2)\big] \nonumber \\
&& - T^a_1 T^a_2 \chi(\bb_1) \chi(\bb_2) -  T^a_{1'} T^a_{2'} \chi(\bb'_1) \chi(\bb'_2) \nonumber \\
&& - T^a_2 T^a_{1'} \chi(\bb_2) \chi(\bb'_1) -  T^a_1 T^a_{2'} \chi(\bb_1) \chi(\bb'_2) \nonumber \\
&& - T^a_1 T^a_{1'} \chi(\bb_1) \chi(\bb'_1) -  T^a_2 T^a_{2'} \chi(\bb_2) \chi(\bb'_2) \, .
\label{eq:4.3.2}\eea
Here products like $ T^a_1 T^a_{2'}$ are shorthands for
$ T^a \otimes 1 \otimes 1 \otimes  T^a $, what acts in the space spanned by
gluon states $g^a_1 \otimes g^b_2 \otimes g^c_{1'} \otimes g^d_{2'}$.

In the following we shall heavily exploit the color--singlet condition
for the four gluon system, which reads
\bea
 \big( T^a_1+ T^a_2+T^a_{1'}+ T^a_{2'}\big)\ket{R\overline{R}} = 0 \, .
\label{eq:4.3.3}\eea
\label{singlet}
As we work with states in which the dipole $12$ and the conjugate dipole
$1'2'$ are in definite color representations, we have
\bea
( T^a_1+ T^a_2 )^2 \ket{R\overline{R}} = (T^a_{1'}+ T^a_{2'})^2\ket{R\overline{R}}
= C_2[R]\ket{R\overline{R}} \, ,
\label{eq:4.3.4}\eea
which we can use to simplify the cross product terms, e.g.
\bea
T^a_1 T^a_2 = T^a_{1'} T^a_{2'} = {1 \over 2} ( C_2[R] - 2 C_A ) \, ,
\label{eq:4.3.5}\eea
where we used that $T^a_jT^a_j = C_A$ for all $j$ 
\footnote{Here a sum over $a$, but not $j$ is implied}.
Now notice that for the operators
\bea
T_D^a &\equiv& T^a_1+ T^a_2 = T^a \otimes 1 \otimes 1 \otimes 1 + 
1 \otimes T^a \otimes 1 \otimes 1 \nonumber \\
T_{D'}^a &\equiv&  T^a_{1'}+ T^a_{2'} = 1 \otimes 1 \otimes T^a \otimes 1 +  
1 \otimes 1 \otimes 1 \otimes T^a \, \nonumber \\
\Delta_D^a  &\equiv& T^a_1 - T^a_2 =  T^a \otimes 1 \otimes 1 \otimes 1 - 
1 \otimes T^a \otimes 1 \otimes 1 \nonumber \\
\Delta_{D'}^a &\equiv&  T^a_{1'} - T^a_{2'} = 1 \otimes 1 \otimes T^a \otimes 1 -  
1 \otimes 1 \otimes 1 \otimes T^a
\label{eq:4.3.6}\eea
we have 
\bea
T_D^a \Delta_D^a = (T_1^a)^2 - (T_2^a)^2 = 0 = T_{D'}^a \Delta_{D'}^a \, ,
\label{eq:4.3.7}\eea
and, because of the color singlet condition $T_D^a = - T_{D'}^a$, also
\bea
T_{D'}^a \Delta_D^a = T_{D}^a \Delta_{D'}^a = 0\, ,
\label{eq:4.3.8}\eea
and, effectively
\bea
T_D^a T_{D'}^a  = - T_D^aT_D^a = - C_2[R]
\label{eq:4.3.9}\eea
Now insert 
\bea
T_1^a = {1 \over 2} ( T_D^a + \Delta_D^a) \, \, ; T_2^a = {1 \over 2} ( T_D^a - \Delta_D^a) \, \, ; 
T_{1'}^a = {1 \over 2} ( T_{D'}^a + \Delta_{D'}^a) \, \, ; T_{2'}^a = 
{1 \over 2} ( T_{D'}^a - \Delta_{D'}^a) \, ,
\label{eq:4.3.10}\eea
into (\ref{eq:4.3.2}), and use relations (\ref{eq:4.3.5}) and (\ref{eq:4.3.10})  to obtain
\bea
1 &-& \textsf{S}_N^{(4)}(\bb_1,\bb_2,\bb'_1,\bb'_2) = 
{1 \over 2} C_A \Big( \chi^2(\bb_1) +\chi^2(\bb_2)+\chi^2(\bb'_1)+\chi^2(\bb'_2)\Big)
\nonumber \\
&&+ {1 \over 2} ( C_2[R] - 2 C_A ) \Big(  \chi(\bb_1) \chi(\bb_2) + \chi(\bb'_1) \chi(\bb'_2) \Big) \nonumber \\
&& - {1 \over 4}  C_2[R] \Big(  \chi(\bb'_1) \chi(\bb_2) +\chi(\bb_1) \chi(\bb'_2) + 
\chi(\bb_1) \chi(\bb'_1) + \chi(\bb_2) \chi(\bb'_2) \Big) \nonumber \\
&& + {1 \over 4 }  \Delta_{D}^a\Delta_{D'}^a \Big( \chi(\bb_1) \chi(\bb'_1) 
+  \chi(\bb_2) \chi(\bb'_2)
-  \chi(\bb'_1) \chi(\bb_2) - \chi(\bb_1) \chi(\bb'_2) \Big) \, .
\label{eq:4.3.11}
\eea
We can now go ahead and complete the squares, to obtain
\bea
1 - \textsf{S}_N^{(4)}(\bb_1,\bb_2,\bb'_1,\bb'_2) &=& {1 \over 4} ( 2 C_A - C_2[R] ) 
\Bigg[ \Big(\chi(\bb_1) - \chi(\bb_2)\Big)^2 + \Big( \chi(\bb_1') - \chi(\bb_2')\Big)^2 \Bigg] \nonumber  \\
&&+ {1 \over 8} C_2[R]  \Bigg[ \Big(\chi(\bb_1) - \chi(\bb'_1)\Big)^2 + 
\Big( \chi(\bb_2) - \chi(\bb_2')\Big)^2 \nonumber \\
&& + \Big(\chi(\bb'_1) - \chi(\bb_2)\Big)^2 + \Big( \chi(\bb_1) - \chi(\bb_2')\Big)^2 \Bigg]
\nonumber \\
&& + {1 \over 8}\Delta_{D}^a\Delta_{D'}^a \Bigg[ \Big(\chi(\bb'_1) - \chi(\bb_2)\Big)^2 + 
\Big( \chi(\bb_1) - \chi(\bb_2')\Big)^2 \nonumber \\
&& - \Big(\chi(\bb_1) - \chi(\bb'_1)\Big)^2 - \Big( \chi(\bb_2) - \chi(\bb_2')\Big)^2 \Bigg] . \label{eq:4.3.12}
\eea

Finally, using (\ref{eq:3.1.3}) 
we obtain the following form for the dipole cross section operator
for the four-gluon system \bea
&&\bra{R'\overline{R'}}\hat{\sigma}^{(4)}(\bb_1,\bb_2,\bb'_1,\bb'_2) \ket{R\overline{R}} 
= \lambda_R \delta_{R',R} \cdot  \Big[ \sigma(\bb_1-\bb_2) + \sigma(\bb'_1 - \bb'_2) \Big] \nonumber \\
&& + { (1-\lambda_R)  \over 2}  \delta_{R',R}  \cdot 
\Big[ \sigma(\bb_1-\bb'_1) + \sigma(\bb_2 - \bb'_2) + \sigma(\bb'_1-\bb_2) + 
\sigma(\bb_1 - \bb'_2) \Big]  \nonumber \\
&& - {\bra{R'\overline{R'}}\Delta_{D}^a\Delta_{D'}^a\ket{R\overline{R}} \over 4 C_A} \cdot 
\Omega(\bb_1,\bb_2,\bb'_1,\bb'_2)
\label{eq:4.3.13}
\eea
Here the parameter
\bea
\lambda_R = 1 - {C_2[R] \over 2 C_A} \, ,
\label{eq:4.3.14}\eea
enters the diagonal part of the cross section and can be found for
individual multiplets in table I. 
In the off-diagonal piece we introduced the combination
\bea
\Omega(\bb_1,\bb_2,\bb'_1,\bb'_2) & \equiv & \sigma(\bb'_1-\bb_2) + \sigma(\bb_1 - \bb'_2) - 
\sigma(\bb_1-\bb'_1) - \sigma(\bb_2 - \bb'_2) \nonumber\\
&=&\sigma(\bs+\br) + \sigma(\bs-\br') - 
\sigma(\bs) - \sigma(\bs+\br-\br')  \, ,
\label{eq:4.3.15}\eea
the same structure of dipole cross sections made already an appearance in
our previous solutions of the $q\bar{q}q\bar{q}$ and $ qg\bar{q}g$
dipole cross section matrices \cite{Nonlinear,QuarkGluonDijet}.
Eq.(\ref{eq:4.3.13}) is the central result of this subsection. 
We now turn to the evaluation of the matrix elements 
${\bra{R'\overline{R'}}}\Delta_{D}^a\Delta_{D'}^a\ket{R\overline{R}}$.


\subsection{Evaluation of the off--diagonal matrix elements}

\begin{figure}[!t]
\begin{center}
\includegraphics[width = 6 cm,height= 8 cm, angle=270]{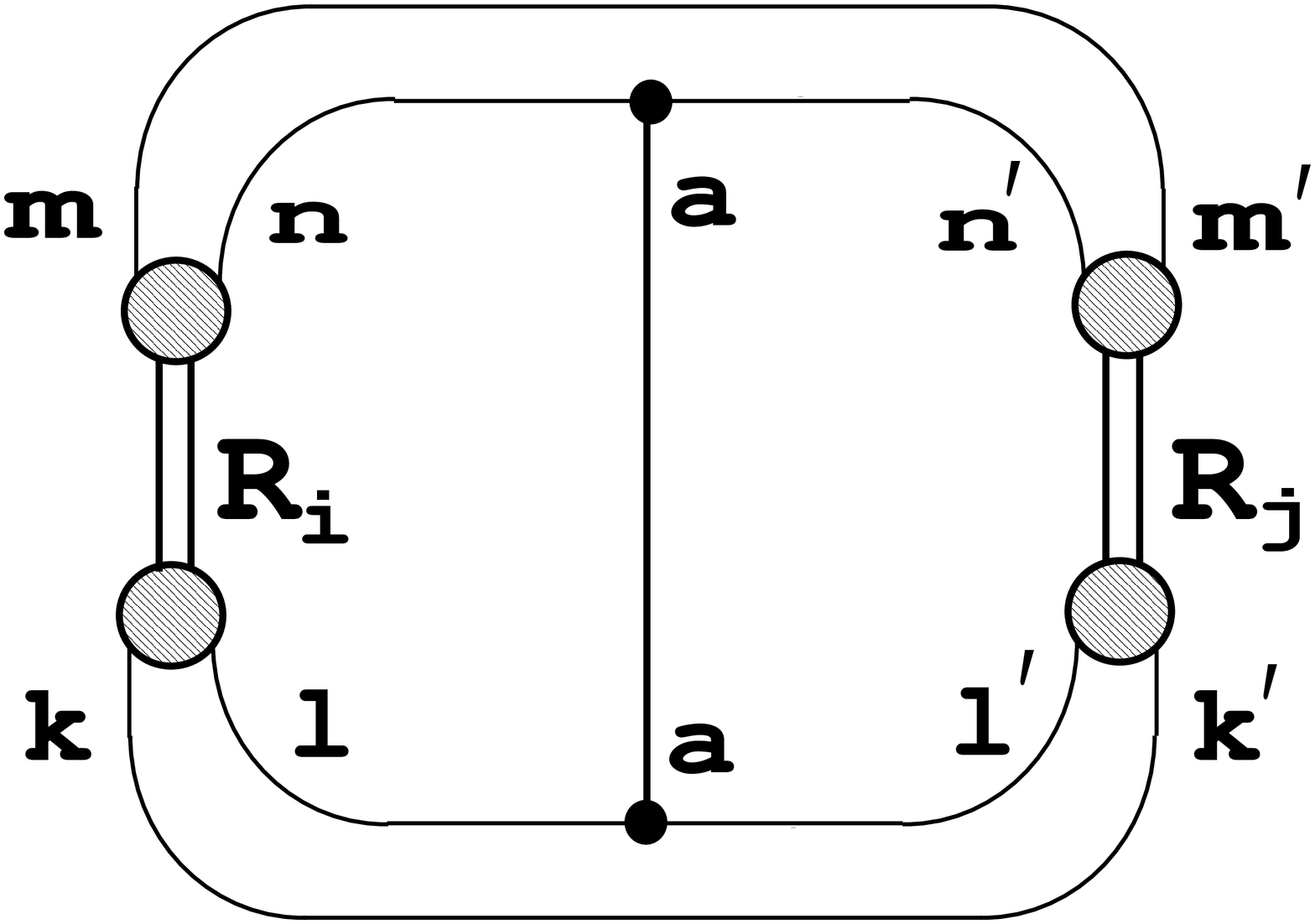}
\caption{The matrix element $\bra{R_j\overline{R_j}}\hat{O} \ket{R_i\overline{R}_i}$
.}
\label{fig:Operator}
\end{center}
\end{figure}

For $R \neq R'$, we have
\bea
\bra{R'\overline{R'}}\hat{\sigma}^{(4)}(\bb_1,\bb_2,\bb'_1,\bb'_2) \ket{R\overline{R}} &=&  
-{\bra{R'\overline{R'}}\Delta_{D}^a\Delta_{D'}^a\ket{R\overline{R}} \over 4 C_A} \Omega(\bb_1,\bb_2,\bb'_1,\bb'_2) 
\label{eq:4.4.1}\eea
We recall, that
\bea
&&\Delta_{D}^a\Delta_{D'}^a = \Big(  T^a \otimes 1 - 
1 \otimes T^a \Big) \otimes  \Big(  T^a \otimes 1 - 1 \otimes T^a \Big) \nonumber \\
&&=  T^a \otimes 1 \otimes T^a \otimes 1 -   1 \otimes T^a \otimes  T^a \otimes 1 -   
T^a \otimes  1 \otimes
 1 \otimes  T^a +  1 \otimes T^a \otimes 1 \otimes T^a \, .
\label{eq:4.4.2}\eea
One can easily convince oneself that this is really a transition operator, i.e., it has only
matrix elements between different multiplets, and even more they must be of different permutation
symmetry, i.e. the transitions are between symmetric and antisymmetric multiplets.
One can then use permutation symmetry, to obtain, effectively,
\bea
\Delta_{D}^a\Delta_{D'}^a &&= 4 \cdot \Big(  1 \otimes T^a \otimes 1 \otimes T^a \Big) \nonumber \\
&&= - 4 \cdot \Big(  1 \otimes T^a \otimes 1 \otimes (T^a)^t \Big) \equiv 4 \cdot \hat{O} \, ,
\label{eq:4.4.3}\eea
which  means
\bea
\bra{R'\overline{R'}}\hat{\sigma}^{(4)}(\bb_1,\bb_2,\bb'_1,\bb'_2) \ket{R\overline{R}} = - {1 \over N_c} \cdot  \bra{R'\overline{R'}}\hat{O} \ket{R\overline{R}}  \cdot \Omega(\bb_1, \bb_2, \bb'_1,\bb'_2) \, .
\label{eq:4.4.4}\eea
Between four gluon states with color wavefunctions $A^{mn}_{jk},B^{mn}_{jk}$ the matrix element 
of the operator ${\cal{O}}$ is evaluated explicitly as
\bea
\bra{A} {\cal{O}} \ket{B} = A^{mn}_{kl} \, i f_{nan'} i f_{all'} \, \delta_{mm'} \delta_{nn'}
B^{k'l'}_{m'n'} \, ,
\eea
see also Fig.(\ref{fig:Operator}).

In table II we collect the contractions of the operator $\hat{O}$ between 
a convenient choice of tensors, from which the matrix elements of the relevant
four gluon states (\ref{eq:4.2.2}) can be reconstructed.
The tensor $iY$ decouples completely -- all its off-diagonal elements vanish.

\begin{table}
\caption{Matrix elements of the Operator $\hat{O}$}
\begin{ruledtabular}
\begin{tabular}{cccccc}
& ${\cal{S}}$ & $P[1]$ & $P[8_S]$ & $ D_u$ & $D_t$ \\
${\cal{A}}$ & ${1 \over 4} N_c(N_c^2-1)^2$ & $N_c$ & ${3\over 4} N_c(N_c^2-1)$ & ${1 \over 2} (N_c^2-1)(N_c^2-4) $ 
& $-{1 \over 4} (N_c^2-1)(N_c^2-4)$ \\
$P[8_A]$ & ${3 \over 4} N_c (N_c^2-1)$ & $N_c$ & ${1 \over 4}N_c(N_c^2-1) $ & $0$ & ${1 \over 4} (N_c^2-1)(N_c^2-4)$   \\
$i Y$ & $0$ & $0 $& $0 $& $0 $& $0$
\end{tabular}
\end{ruledtabular}
\end{table}


\subsection{The dipole cross section operator for the four--gluon system}

We now come to the final result for the four-body  dipole cross section 
operator in the $s$-channel basis described in Sec. IV B above, in
which the two gluons from the amplitude and from the complex conjugate amplitude,
respectively, are in definite color multiplets.


\subsubsection{Diagonal elements}

The diagonal elements for the four-gluon system are even
simpler than for the quark-antiquark-gluon-gluon
system studied in \cite{QuarkGluonDijet} and are expressed in terms of two combinations of
color-dipole cross sections:
\bea\Sigma_1& \equiv &
\sigma(\bb_1 - \bb_2) + \sigma(\bb_1'-\bb_2') =\sigma(\br)+\sigma(-\br')\, ,\nonumber\\
\tau &\equiv& \sigma(\bb_1-\bb'_1) + \sigma(\bb_2-\bb'_2) +
\sigma(\bb'_1-\bb_2) + \sigma(\bb_1 -\bb'_2) \nonumber\\
&=&\sigma(\bs+\br)+\sigma(\bs-\br') +\sigma(\bs)+\sigma(\bs+\br-\br')\, . 
 \label{eq:4.5.1.1}\eea
Making use of table I, we find
\bea
\bra{11} \hat{\sigma}^{(4)} \ket{11} &=& \Sigma_1 \nonumber \\
\bra{8_A8_A} \hat{\sigma}^{(4)} \ket{8_A8_A} &=& 
\bra{8_S8_S} \hat{\sigma}^{(4)} \ket{8_S8_S}= { 1\over 4}  \tau +{1\over 2}\Sigma_1   \nonumber \\
\bra{10\overline{10}} \hat{\sigma}^{(4)} \ket{10 \overline{10}}  
&=& \bra{\overline{10}10} \hat{\sigma}^{(4)} \ket{\overline{10}10}
={1 \over 2} \tau = {C_2[10]\over C_A} \cdot { 1\over 4}  \tau \nonumber \\
\bra{2727} \hat{\sigma}^{(4)} \ket{2727} &=&  
{2(N_c+1) \over N_c}\cdot  {1 \over 4} \tau -{1\over 2N_c}\Sigma_1 =
{C_2[27]\over C_A} \cdot { 1\over 4}  \tau -
{1\over 2N_c}\Sigma_1
\nonumber \\
\bra{R_7R_7} \hat{\sigma}^{(4)} \ket{R_7R_7} &=& 
{2(N_c-1) \over N_c}\cdot  {1 \over 4} \tau +{1\over 2N_c}\Sigma_1 =
{C_2[R_7]\over C_A} \cdot { 1\over 4}  \tau +
{1\over 2N_c}\Sigma_1
\label{eq:4.5.1.2}\eea
We recall that in the limit of $\br=\br'=0$ the gluon pairs collapse into
pointlike partons in the color representation $R_i$. In this limit $\Sigma_1=0$,
and the diagonal matrix elements are simply proportional to the Casimir
operators as it must be \cite{QuarkGluonDijet}:
\bea
\bra{R_i R_i} \hat{\sigma}^{(4)} \ket{R_i R_i} 
 ={C_2[R_i]\over C_A} \sigma(\bs).
\label{eq:4.5.1.3}\eea
Notice that the matrix elements $\bra{2727} \hat{\sigma}^{(4)} \ket{2727}$ and 
$\bra{R_7R_7} \hat{\sigma}^{(4)} \ket{R_7R_7}$ are related by the transformation $N_c\to -N_c$,
for the discussion of a similar symmetry in the quark-gluon dijet production see
Ref. \cite{QuarkGluonDijet}  

\subsubsection{Off-diagonal elements}

Making use of Tables I,II we have

\bea
\bra{11} \hat{\sigma}^{(4)} \ket{8_A8_A} &=& 
-{1 \over \sqrt{N_c^2-1}} \cdot  \Omega(\bb_1, \bb_2, \bb'_1,\bb'_2) \nonumber \\
\bra{8_S8_S} \hat{\sigma}^{(4)} \ket{8_A8_A} &=& 
- {1 \over 4} \cdot  \Omega(\bb_1, \bb_2, \bb'_1,\bb'_2) \nonumber \\
\bra{2727} \hat{\sigma}^{(4)} \ket{8_A8_A} &=& 
-{1 \over 2N_c} \sqrt{N_c+3 \over N_c+1} \cdot  \Omega(\bb_1, \bb_2, \bb'_1,\bb'_2) \nonumber \\
\bra{R_7R_7} \hat{\sigma}^{(4)} \ket{8_A8_A} &=& 
-{1 \over 2N_c} \sqrt{N_c-3 \over N_c-1} \cdot  \Omega(\bb_1, \bb_2, \bb'_1,\bb'_2) \nonumber \\
\bra{8_S8_S} \hat{\sigma}^{(4)} \ket{10\overline{10}} &=& 
- {1 \over 2\sqrt{N_c^2-4}} \cdot  \Omega(\bb_1, \bb_2, \bb'_1,\bb'_2) \nonumber \\
&=& \bra{8_S8_S} \hat{\sigma}^{(4)} \ket{\overline{10}10} \nonumber \\
\bra{2727} \hat{\sigma}^{(4)} \ket{10\overline{10}}  &=& 
-{1 \over 4N_c} \sqrt{{(N_c+1)(N_c-2)(N_c+3)\over N_c+2}} \cdot  \Omega(\bb_1, \bb_2, \bb'_1,\bb'_2) \nonumber \\
&=& \bra{2727} \hat{\sigma}^{(4)} \ket{\overline{10}10} \nonumber \\
\bra{R_7R_7} \hat{\sigma}^{(4)} \ket{10\overline{10}}  &=&
- {1 \over 4N_c} \sqrt{{(N_c-1)(N_c+2)(N_c-3)\over N_c-2}} \cdot  \Omega(\bb_1, \bb_2, \bb'_1,\bb'_2) \nonumber \\
&=& \bra{R_7R_7} \hat{\sigma}^{(4)} \ket{\overline{10}10} \nonumber \\
\label{eq:4.5.2.1}\eea
We again observe the curious symmetry \cite{QuarkGluonDijet}: the matrix elements
$\bra{2727} \hat{\sigma}^{(4)} \ket{8_A8_A}$ and $\bra{R_7R_7} \hat{\sigma}^{(4)} \ket{8_A8_A}$
are related by the transformation  $N_c\to -N_c$, the same is true of the matrix elements
 $\bra{2727} \hat{\sigma}^{(4)} \ket{\overline{10}10}$ and $\bra{R_7R_7} \hat{\sigma}^{(4)} \ket{10\overline{10}}$.


\subsection{Large-$N_c$ properties of the dipole cross section
matrix}

In conjunction with the Glauber--Gribov form for the nuclear
$S$--matrix, the dipole cross section operator
$\hat{\sigma}^{(4)}$ solves the problem of non-Abelian intranuclear
evolution of the four--gluon system.  Being a symmetric matrix,
the dipole cross section operator could readily be brought in
diagonal form, and the Fourier-transform could finally be
performed numerically. In practice, however one would encounter a
number of obstacles when proceeding along these lines. First, for
the case of the two-particle inclusive spectrum, the eigenvalues
of the dipole cross section operator will be non-algebraic
functionals of the free-nucleon cross section operator, and second
the dipole cross section itself has a non-analytic dependence on
dipole size, which would ultimately determine the asymptotics of
the Fourier-transforms. It is therefore convenient that the
large-$N_c$ expansion offers a path to analytic formulas, which
can be interpreted in a transparent way.

We start from the observation, that at large-$N_c$ the cross
section operator $\hat{\sigma}^{(4)}$ assumes a block diagonal
form. Apparently, transitions between representations which have
dimensions that grow with the same power of $N_c$ are
parametrically of order $N_c^0$, whereas transitions to the
next-larger/smaller block are suppressed by $N_c^{-1}$.

In the space of four parton-states we introduce the projectors
\bea
{\cal{P}}_1 &&= \ket{11}\bra{11} \nonumber \\
{\cal{P}}_2 &&= \ket{8_A8_A}\bra{8_A8_A} + \ket{8_S8_S}\bra{8_S8_S}  \nonumber \\
{\cal{P}}_3 &&= \ket{10\overline{10}}\bra{10\overline{10}} +
\ket{\overline{10}10}\bra{\overline{10}10} + \ket{2727}\bra{2727}
+ \ket{R_7 R_7}\bra{R_7 R_7} \, , \label{eq:4.6.1}\eea
which allow us to isolate the blocks ${\cal{P}}_i \,
\hat{\sigma}^{(4)} \,  {\cal{P}}_i$, the first one being a
one-by-one matrix:
\bea {\cal{P}}_1 \hat{\sigma}^{(4)} {\cal{P}}_1  = \Sigma_1 \, . \label{eq:4.6.2}\eea
The vector $\ket{e_1} =\ket{11}$ can be viewed as an eigenvector of
block $1$ with eigen-cross-section $\Sigma_{1}$.

The second block is written, in the two-dimensional subspace of
octets, as
\bea {\cal{P}}_2 \hat{\sigma}^{(4)} {\cal{P}}_2 = \left (
\begin{array}{cc}
{1 \over 4}[\tau + 2 \Sigma_1] & -{1\over 4} \Omega \\
-{1 \over 4} \Omega & {1 \over 4} [\tau + 2 \Sigma_1]
\end{array} \right ) \, . \label{eq:4.6.3}\eea
Its eigenvectors are
\bea \ket{e_2} && = {1 \over \sqrt{2}} ( \ket{8_A 8_A} + \ket{8_S
8_S} ) \nonumber \\
\ket{e_3} &&={1 \over \sqrt{2}} ( \ket{8_A 8_A} - \ket{8_S 8_S} )
\, , \label{eq:4.6.4}\eea
and belong to the eigen-cross-sections
\bea 
\Sigma_2 & =& {1 \over 2} \Sigma_1 + { 1 \over 4} [\tau - \Omega]
={1 \over 2} \Big( \Sigma_1 + \sigma(\bb_1 - \bb_1') +
\sigma(\bb_2 - \bb_2') \Big)   \nonumber \\
&=& {1\over 2}\Big[\sigma(\br)+\sigma(-\br')+\sigma(\bs)+\sigma(\bs+\br-\br')\Big]\,,
\nonumber\\
\Sigma_3 & =& {1 \over 2} \Sigma_1  + {1 \over 4} [\tau +
\Omega] = { 1\over 2} \Big[ \Sigma_1 + \sigma(\bb_1'-\bb_2)  + \sigma(\bb_1 -\bb_2') \Big] \nonumber \\
&=& {1\over 2}\Big[\sigma(\br)+\sigma(-\br')+\sigma(\bs+\br)+\sigma(\bs-\br')\Big]
\, . \label{eq:4.6.5}\eea
Finally, the third block accounts for the multiplets that have
${\cal{O}}(N_c^4)$ states. Here we notice that at large $N_c$
all higher multiplets interact as two color-uncorrelated gluons.
For instance, the Casimir operators for these multiplets  
approach $C_2(R_i)=2C_A$ and the diagonal cross-sections become
identical. In matrix form, where the rows refer to
states
$\ket{10\overline{10}},\ket{\overline{10}10},\ket{2727},\ket{R_7
R_7}$ we write
\bea {\cal{P}}_3 \hat{\sigma}^{(4)} {\cal{P}}_3 = \left (
\begin{array}{cccc}
{1 \over 2}\tau  & 0 & -{1\over 4} \Omega & -{1 \over 4 }\Omega \\
0 & {1 \over 2} \tau & -{1 \over 4} \Omega &-{1 \over 4} \Omega \\
-{1 \over 4}\Omega & -{1 \over 4}\Omega & {1 \over 2} \tau & 0 \\
-{1 \over 4}\Omega & -{1 \over 4}\Omega &  0 & {1 \over 2} \tau
\end{array} \right ) \, , \label{eq:4.6.6}\eea
Its eigenvectors are easily seen to be
\bea \ket{e_4} &&= {1\over 2} \Big( \ket{10\overline{10}} +
\ket{\overline{10}10} + \ket{2727} + \ket{R_7 R_7} \Big )
\nonumber \\
\ket{e_5} && = {1\over 2} \Big(\ket{10\overline{10}} +
\ket{\overline{10}10} - \ket{2727} - \ket{R_7 R_7} \Big)\nonumber
\\
\ket{e_6} && = {1 \over \sqrt{2}} \Big(\ket{10\overline{10}} -
\ket{\overline{10}10} \Big) \nonumber \\
\ket{e_7} && ={1 \over \sqrt{2}} \Big(  \ket{2727} - \ket{R_7 R_7}
\Big) \, , \label{eq:4.6.7}\eea
with eigenvalues
\bea \Sigma_4  = {1\over 2}( \tau - \Omega) &=& \sigma(\bb_1 -
\bb_1') + \sigma(\bb_2-\bb_2') \nonumber \\&=& \sigma(\bs)+\sigma(\bs+\br-\br')\nonumber\\
\Sigma_5  ={ 1\over 2} ( \tau+\Omega) &=& \sigma(\bb_1'- \bb_2) +
\sigma(\bb_1 - \bb_2') \nonumber \\&=& \sigma(\bs+\br)+\sigma(\bs-\br')\nonumber\\
\Sigma_6  = \Sigma_7 = {1 \over 2} \tau &=& {1 \over 2} ( \Sigma_4
+ \Sigma_5) \, .\label{eq:4.6.8} \eea

We finally observe, that the eigenstate $\ket{e_6}$ decouples
exactly from our problem, which is seen readily from the summary of the
off-diagonal elements (\ref{eq:4.5.2.1}).
Couplings between the above diagonalized matrix blocks are of
${\cal{O}}(N_c^{-1})$. 
In the basis of eigenstates
$\ket{e_1},\dots,\ket{e_7}$ we can collect the $N_c$-suppressed
off--diagonal elements as
\bea \hat{\omega}(\bs,\br,\br') = -{1 \over\sqrt{2} N_c}
\Omega(\bs,\br,\br') \Big\{  \ket{e_1}\bra{e_2} +
\ket{e_1}\bra{e_3} + \ket{e_4}\bra{e_2} - \ket{e_5}\bra{e_3} +
\mathrm{h.c.} \Big\} \, . \label{eq:4.6.9}\eea
It is easy to check that the off-diagonal
elements containing the state $\ket{e_7}$ are ${\cal{O}}(N_c^{-2})$
and this state decouples at ${\cal{O}}(N_c^{-1})$. 
At ${\cal{O}}(N_c^{-1})$ there are also corrections to the
diagonal matrix elements, in the sector of large symmetric
representations $27, R_7$. They are however not relevant for our
problem, the $N_c^{-1}$ perturbation theory treatment
of such corrections is found in \cite{Nonlinear}. 
The crucial observation is, that in the inclusive dijet
spectrum, with summation over all colors of final state gluons,
the final state projection simplifies to
\bea \sum_X \bra{X} = \sum_R \sqrt{\dim[R]} \bra{R\overline{R}} =
\underbrace{\bra{e_1}}_1 + \underbrace{N_c \sqrt{2} \bra{e_2}}_{8_A
+8_S} + \underbrace{ N_c^2
\bra{e_4}}_{10+\overline{10}+27+R_7}\,. \label{eq:4.6.10}\eea
If we remember, that averaging over incoming colors shall
introduce another factor $1/N_c$,
\bea \ket{\mathrm{in}} = {1 \over  \sqrt{\dim[8]}} \ket{8_A 8_A} = {1
\over \sqrt{2}N_c} \Big(\ket{e_2} +\ket{e_3}\Big) \, , \label{eq:4.6.11}\eea
we see that the large number of states in higher multiplets can
overcome the suppression of their excitation. Furthermore, the
excitation of singlet states is large-$N_c$ suppressed and will be
neglected from now on.

To develop the large-$N_c$ perturbation theory, we decompose the
free--nucleon cross section operator as
\bea \hat{\sigma}^{(4)} = \hat{\Sigma}^{(0)} + \hat{\omega} \, ,
\label{eq:4.6.12}\eea
where
\bea \hat{\Sigma}^{(0)} ={\cal{P}}_1 \hat{\sigma}^{(4)}
{\cal{P}}_1 +  {\cal{P}}_2 \hat{\sigma}^{(4)} {\cal{P}}_2 +
{\cal{P}}_3 \hat{\sigma}^{(4)} {\cal{P}}_3 =\sum_{j=1}^7 \Sigma_j\ket{e_j}\bra{e_j}
\label{eq:4.6.13}\eea
is the block matrix that is diagonalized by the basis $\ket{e_1}
\dots \ket{e_7}$. Now, the nuclear $S$--matrix is obtained from
the free nucleon dipole cross section by means of the
Glauber--Gribov exponentiation
\bea \textsf{S}[\bb, \hat{\sigma}^{(4)}(\bs,\br,\br')] = \exp[-{1
\over 2} \hat{\sigma}^{(4)}(\bs,\br,\br') T(\bb)] \, . \label{eq:4.6.14}\eea
To the first order in the off-diagonal perturbation $\hat{\omega}$
we can easily establish
\bea \textsf{S}[\bb, \hat{\Sigma}^{(0)} + \hat{\omega}]
-\textsf{S}[\bb, \hat{\Sigma}^{(0)}] = -{1 \over 2} T(\bb) \int_0^1
d\beta \, \textsf{S}[\bb, (1- \beta)  \hat{\Sigma}^{(0)}] \,
\hat{\omega} \, \textsf{S}[\bb, \beta \hat{\Sigma}^{(0)}] +
{\cal{O}}(N^{-2}). \label{eq:4.6.15}\eea



\section{Linear $k_\perp$-factorization for dijets from the free
nucleon target}

At this point we are in a position to give our result for the
process $g^* g_N \to g_1 g_2$ on the free nucleon target. After
integrating over the overall impact parameter, our master-formula
assumes the form
\bea {d \sigma_N(g^* \to g_1g_2) \over dz d^2\bp_1 d^2\bp_2 } && =
{1 \over 2 (2 \pi)^4} \int d^2\bs d^2\br d^2\br' \exp[-i \bDelta \bs 
- i \bp_1 (\br - \br')] \Psi(z, \br)
\Psi^*(z,\br') \nonumber \\
&&\times \sum_{X}\bra{X} {\hat{\Sigma}}(\bs, \br, \br')
\ket{in}
\label{eq:5.1} \eea

We evaluate the cross section for an incoming gluon, which entails
an  average over its incoming colors and the initial state
$\ket{in}$ of Eq.~(\ref{eq:4.2.1}).
Likewise in the final state we sum over all color states.
Then, the calculation of the inclusive cross section involves the
evaluation of the following matrix elements of the four parton
cross section operator
\bea 
&& \sum_{R \neq 8_A} \sqrt{\dim [R] \over \dim [8]}
\bra{R\overline{R}} {\hat{\Sigma}}(\bs,\br,\br') \ket{8_A
8_A} =\nonumber\\ 
&&= - \sum_{R \neq 8_A} \sqrt{\dim [R] \over N_c^2-1}
\bra{R\overline{R}} {\hat{\sigma}}^{(4)}(\bs,\br,\br') \ket{8_A
8_A} = 
\Omega(\bs,\br,\br') \nonumber \\
&& \times \Biggl( \underbrace{1 \over 4}_{8_S}  + {1 \over  {\underbrace{N_c^2-1}_{1}}} +   
\underbrace{\sqrt{{N_c^2 (N_c+3)(N_c-1) \over
4 (N_c^2-1) }} {1\over 2N_c}\sqrt{{N_c+3 \over N_c+1}}}_{27} \nonumber\\
&&+  \underbrace{\sqrt{{N_c^2
(N_c-3)(N_c+1) \over
4 (N_c^2-1) }} {1\over 2N_c}\sqrt{{N_c-3 \over N_c-1}}}_{R_7} \Biggr) \nonumber \\
&& = {1 \over 4} \Omega(\bs,\br,\br') \Biggl( \underbrace{1}_{8_S}
+ {4 \over {\underbrace{N_c^2-1}_{1}}} + {N_c+3 \over
{\underbrace{N_c+1}_{27}}} + {N_c-3 \over \underbrace{N_c-1}_{R_7} }\Biggr)
\nonumber \\
&& ={1 \over 4} \, \Omega(\bs,\br,\br') \,  \Big(
\underbrace{1}_{8_S} + \underbrace{2}_{1 + 27 + R_7} \Big) \, .
 \label{eq:5.3}\eea
Here we indicated the contributions from excitation of
individual multiplets. $\Omega(\bs,\br,\br')$ is the same as in
eq. (\ref{eq:4.3.15}), now in the relevant coordinates, explicitly
\bea \Omega(\bs,\br ,\br')  = \int d^2\bkappa f(\bkappa) \, \exp[i \bkappa \bs] 
(1 - \exp[i\bkappa\br]) (1-\exp[-i\bkappa \br'])  \, .  \label{eq:5.4}\eea
The diagonal contribution from final state gluons in the
antisymmetric octet is easily constructed from the results of
sec. IV:
\bea && \bra{8_A8_A} {\hat{\Sigma}}(\bs,\br ,\br') \ket {8_A 8_A} =
\sigma^{(3)}(\bb_1,\bb_2,\bb') + \sigma^{(3)}(\bb'_1,\bb'_2,\bb)
- \sigma(\bs + z (\br - \br')) \nonumber \\
&&  - \bra{8_A8_A} {\hat{\sigma}}^{(4)}(\bs,\br ,\br') \ket {8_A 8_A} \nonumber \\
&& = {1 \over 2} \Big[ \sigma(\bs -z\br'+\br) +
\sigma(\bs-\br'+z\br ) - \sigma(\bs +z(\br -\br') ) - \sigma(\bs +
\br -\br') \nonumber \\
&& + \sigma(\bs -z \br') + \sigma(\bs + z \br) - \sigma(\bs +z(\br
-\br') ) - \sigma(\bs ) \Big] \nonumber \\
&& - {1 \over 4} \Big[ \sigma(\bs -\br') + \sigma(\bs +\br) 
- \sigma(\bs + \br -\br') - \sigma(\bs) \Big] \nonumber \\
&& =  \int d^2 \bkappa f(\bkappa) \exp[i\bkappa \bs]
\nonumber \\
&&\times \Bigg\{  { 1\over 2}  \Big(\exp[i\bkappa \br] - \exp[i z
\bkappa\br]\Big) \Big(\exp[-i\bkappa\br'] - \exp[-iz\bkappa \br']\Big)
\nonumber \\
&&+ {1 \over 2} \Big(1 -\exp[i z \bkappa\br]\Big) \Big(1 - \exp[-i z\bkappa \br']\Big)\nonumber\\
&& - {1 \over 4} \Big(1 -\exp[i  \bkappa\br]\Big) \Big(1 - \exp[-i \bkappa
\br']\Big) \Bigg\} \, .  \label{eq:5.5}\eea
The dijet cross section for production of gluons in the
antisymmetric octet reads therefore
\bea {d \sigma_N(g^* \to \{g_1g_2\}_{8_A} ) \over dz d^2\bDelta
d^2\bp }  = {1 \over 2 (2\pi)^2} f(\bDelta) &\Big\{& |\Psi(z, \bp
-\bDelta) - \Psi(z,\bp-z\bDelta)|^2 + \nonumber\\
&+&|\Psi(z,\bp) - \Psi(z,
\bp-z\bDelta)|^2 \nonumber \\
&-& {1\over 2}|\Psi(z,\bp) - \Psi(z,\bp-\bDelta) |^2 \Big\} \, ,
 \label{eq:5.6}\eea
The wave function of the gluon-gluon Fock state of the physical gluon
enters our analysis as the recurrent quantity 
\bea
|\Psi(z,\bp)-\Psi(z,\bp-\bkappa)|^2 =2P_{gg}(z)
\left({\bp \over  \bp^{2}+\varepsilon^{2}} -
{\bp-\bkappa  \over  (\bp-\bkappa )^{2}+\varepsilon^{2}}\right)^2
, 
\label{eq:5.7}
\eea
where $P_{gg}(z)$ is the familiar gluon splitting function,
\bea
P_{gg}(z)&=&2C_A \left[{1-z \over z} + {z \over 1-z } + z(1-z) \right],\nonumber\\
\varepsilon^2&=& z(1-z) (Q^*)^2.
\label{eq:5.8}
\eea

Now notice, that the last contribution in (\ref{eq:5.6}) 
would be canceled exactly,
in the fully inclusive sum, by the excitation of symmetric octets.
Summing over all possible final states we end up with
\bea {d \sigma_N(g^* \to  g_1g_2 ) \over dz d^2\bDelta d^2\bp }
 = {1 \over 2 (2\pi)^2} f(\bDelta) &\Big\{& |\Psi(z, \bp -\bDelta)
-\Psi(z,\bp-z\bDelta)|^2 \nonumber\\ &+&  |\Psi(z,\bp) - \Psi(z,
\bp-z\bDelta)|^2 \nonumber \\
& +& |\Psi(z,\bp) - \Psi(z,\bp-\bDelta) |^2 \Big\} \, ,
\label{eq:5.9} \eea
where now the last line sums up the contribution from excitation
of $1, 27, R_7$, whereas the first two terms represent the sum of
octet, $8_A,8_S$ final states. No large $N_c$ approximation has
been invoked here,  
what would change  with $N_c$ is only the composition of the final
state, where the excitation of color singlet states is
${\cal{O}}(N_c^{-2})$. The absence of decuplet excitation is a
consequence of the excitation mechanism being single--gluon
exchange. Our result eq. (\ref{eq:5.9}) is of course nothing
but the differential version of eq. (82) of Ref.\cite{SingleJet},
further elucidating the color-composition of the final state. By
itself it would find interesting applications to dijet production
in a regime where saturation/absorption effects are not strong,
e.g. central dijets at HERA or Tevatron.


\section{Nonlinear $k_\perp$-factorization for dijets from nuclear
targets}


\subsection{Dijets in color-octet final states}

In order to evaluate our master formula for the case of octet-final states we 
have to collect the various multiparton $S$--matrices. The three-- and
two--body $S$--matrices are single channel problems and can be written
in terms of the Glauber--Gribov exponential as
\bea
\textsf{S}^{(3)}_{g'_1 g'_2 g^*} && = \textsf{S}[\bb, \sigma_{g'_1 g'_2 g^*}] 
 = \textsf{S} [\bb, \T \half \sigma(\br)] \, \textsf{S}[\bb,\half \sigma(\bs - z \br' + \br) ] 
\, \textsf{S}[\bb, \half \sigma(\bs - z \br')] \nonumber \\
\textsf{S}^{(3)}_{g_1 g_2 g'^*} && = \textsf{S}[\bb, \sigma_{g_1 g_2 g'^*}] 
 = \textsf{S} [\bb, \T \half \sigma(\br')]\, \textsf{S}[\bb,\half \sigma(\bs - \br' + z \br) ] 
\, \textsf{S}[\bb, \half \sigma(\bs + z \br)] \nonumber \\
\textsf{S}^{(2)}_{g^* g'^*} && = \textsf{S}[\bb, \sigma_{g^*g'^*}] 
 = \textsf{S} [\bb, \sigma(\bs + z(\br -  \br'))] =  \textsf{S}^2 [\bb, \T \half 
\sigma(\bs + z(\br -  \br'))]
\label{eq:6.1.1}\eea
Notice that the factorization of $S$--matrices is an exact consequence 
of the form of the dipole cross section for the three-gluon state and 
of the Glauber--Gribov exponentiation valid for a large nucleus, 
it is not related to the large--$N_c$ limit.
We need to invoke the large--$N_c$ approximation only for the contribution
from the four--body $S$--matrix
\bea
\bra{e_2} \textsf{S}[\bb,\hat{\Sigma}^{(0)}] \ket{e_2} =  \textsf{S}[\bb,\Sigma_2] =
\textsf{S}[\bb,\T{1 \over 2}\sigma(\br)] \, \textsf{S}[\bb, \half \sigma(\br')] \, 
\textsf{S}[\bb, \half \sigma(\bs)] \, \textsf{S}[\bb, \half \sigma(\bs + \br - \br')] \, .
\label{eq:6.1.2}\eea

Our aim is to find a momentum space representation in terms of the nuclear unintegrated
glue. The latter is given by the pertinent function $\phi_g$ defined through \cite{SingleJet}
\bea
 1 - \textsf{S}[\bb,\half \sigma(\br)] \equiv \int d^2\bkappa \phi_g(\bb,\bkappa) 
(1 - \exp[i \bkappa\br]) \, .
\label{eq:6.1.3}\eea
Notice a subtlety:
\bea 
\textsf{S}[\bb,\textstyle \half \sigma(\br)] = \exp[- {1 \over 4} \sigma(\br) T(\bb)] =
\exp[- \displaystyle {C_A \over 4C_F} \sigma_{q\bar{q}}(\br) T(\bb)],
\label{eq:6.1.4}\eea
where we used the relationship between the dipole cross section $\sigma(\br)$ defines
for a gluon-gluon system and the quark-antiquark system,
\bea
\sigma(\br ) = {C_A \over C_F} \sigma_{q\bar{q}}(\br).
\label{eq:6.1.5}\eea
Consequently, the collective unintegrated nuclear glue  $\phi_g(\bb,\bkappa)$ is different form the nuclear 
unintegrated glue $\phi(\bb,\bkappa)$ that enters deep inelastic scattering as well as diffractive quark--antiquark
jet production and has been introduced in \cite{Nonlinear,NSSdijet}.
It should not be mixed up with yet another quantity, the unintegrated glue
defined through the color--singlet gluon--gluon probe, $\phi_{gg}(\bkappa)$, 
 which is defined through (for more discussion see Appendix D)
\bea
 1 - \textsf{S}[\bb,\sigma(\br)] \equiv \int d^2\bkappa \phi_{gg}(\bb,\bkappa) (1 - \exp[i \bkappa\br]) \, .
\label{eq:6.1.6}\eea

If we denote by 
\bea
\sigma_0 \equiv {C_A \over C_F } \int d^2\bkappa f (\bkappa) \, 
\label{eq:6.1.7}\eea
the dipole cross section for a large gluon--gluon dipole, then
$\phi_{gg},\phi_g$ are related as
\bea
\phi_{gg}(\bb,\bkappa) &&= 2 \textsf{S}[\bb,\half \sigma_0] \phi_g(\bb,\bkappa) + 
\Big(\phi_g \otimes \phi_g \Big) (\bb,\bkappa) \nonumber \\
 \Big(\phi_g \otimes \phi_g \Big) (\bb,\bkappa) && 
= \int d^2\bq \phi_g(\bb, \bkappa-\bq) \phi_g(\bb, \bq) \, .
\label{eq:6.1.8}
\eea

Another useful quantity is 
\bea
\Phi_g(\bb,\bkappa) = \textsf{S}[\bb,\T {1 \over 2} \sigma_0] \delta^{(2)}(\bkappa)   + \phi_g(\bb,\bkappa) \, ,
\label{eq:6.1.9}\eea
in terms of which,
\bea
\textsf{S}[\bb, \sigma(\br)] = \int d^2 \bkappa \, \Phi_g(\bb, \bkappa) \, \exp[i \bkappa \br] \, .
\label{eq:6.1.10}\eea

For later applications we shall also use the collective glue for a slice
$0 < \beta < 1$ of a nucleus
\bea
 \textsf{S}[\bb,\beta \sigma(\br)] = \int d^2 \bkappa \, \Phi_g(\beta; \bb, \bkappa) \, \exp[i \bkappa \br] \, .
\label{eq:6.1.11}\eea
It has the convolution property
$\Big(\Phi_g(\beta_1)\otimes \Phi_g(\beta_2)\Big)(\bb, \bkappa)=\Phi_g(\beta_1+\beta_2; \bb, \bkappa)$ and
we note, that eq.~(\ref{eq:6.1.8}) amounts to
$ \Phi_{gg}(\bb, \bkappa)= \Phi_{g}(2,\bb, \bkappa)$.

Intranuclear attenuation of dipoles becomes manifest in the distorted wave functions,
defined in dipole and momentum space, respectively, as
\bea
\Psi(\beta;z,\br) &\equiv&  \textsf{S}[\bb,\beta \sigma(\br)] \, \Psi(z,\br) \nonumber \\
\Psi(\beta;z,\bp) &=& \int d^2\br \exp[-i \bp\br] \Psi(\beta;z,\br) = 
\int d^2\bkappa \Phi_g(\bb,\bp-\bkappa) \Psi(z, \bp) \, .
\label{eq:6.1.12}\eea
Now, using our master formula, the inclusive dijet cross section for gluons
in the octet final state unfolds as
\bea
&&{d \sigma (g^* \to \{g_1g_2 \}_{8_A + 8_S} ) \over d^2\bb dz d^2\bp d^2\bDelta } = 
{1 \over (2 \pi)^4} \int d^2\bs d^2\br d^2\br' \exp[-i \bDelta \bs] \exp[-i \bp (\br - \br')] \nonumber \\
&& \Big\{ \Psi(1;z,\br) \Psi^*(1;z,\br') \, \textsf{S}[\bb,\T \half  \sigma(\bs + \br - \br')]  
\textsf{S}[\bb,\half  \sigma(\bs )] \nonumber \\
&& + \Psi(z,\br) \Psi^*(z,\br') \, \textsf{S}^2 [\bb, \T \half 
\sigma(\bs + z(\br -  \br'))] \nonumber \\
&& - \Psi(1;z,\br) \Psi^*(z,\br') \,  \textsf{S}[\bb, \T \half \sigma(\bs - z \br' + \br) ] 
\, \textsf{S}[\bb, \half \sigma(\bs - z \br')]  \nonumber \\
&& - \Psi(z,\br) \Psi^*(1;z,\br') \,  \textsf{S}[\bb,\T \half \sigma(\bs - \br' + z \br) ] 
\, \textsf{S}[\bb, \half \sigma(\bs + z \br)] \Big\} \nonumber \\
&& = {1 \over 2(2 \pi)^2} \int d^2 \bkappa_1 \int d^2 \bkappa_2 \delta(\bDelta- \bkappa_1 - \bkappa_2)
\Phi_g(\bb,\bkappa_2) \Phi_g(\bb,\bkappa_1) \nonumber\\ &\times&
\Big\{ |\Psi(1;z,\bp-\bkappa_1) - \Psi(z, \bp - z \bkappa_1 - z \bkappa_2)|^2 +
|\Psi(1;z,\bp-\bkappa_2) - \Psi(z, \bp - z \bkappa_1 - z \bkappa_2)|^2
\Big\} \nonumber \\
\label{eq:6.1.13}
\eea
Here we presented the result in a manifestly $\bkappa_1\leftrightarrow \bkappa_2$ 
symmetric form.
This result fully conforms with the concept of universality classes introduced in Refs. 
\cite{Nonuniversality,QuarkGluonDijet} and must be compared to the dijet spectra of other
reactions in which the dijets are produced in the same color representation as
the incident parton: excitation of color-octet quark-antiquark 
dijet or of open heavy flavor from gluons, $g\to \{Q\bar{Q}\}_8$, and
excitation of color-triplet quark-gluon dijets from quarks, $q\to \{qg\}_3$. 
The incoherent distortion factor 
$\Phi_g(\bb,\bkappa_1) \Phi_g(\bb,\bkappa_2)$ in the integrand is the same
as in  another gluon induced reaction $g\to Q\bar{Q}$. It must be compared
to the distortion factor $\Phi_g(\bb,\bDelta)$ in the case of  $q\to  \{qg\}_3$.
Following \cite{Nonuniversality,QuarkGluonDijet} we note, that (i) to 
the considered large-$N_c$ approximation the above $\Phi_g(\bb,\bkappa)$
equals the nuclear collective glue 
$\Phi(\bb,\bkappa)$ defined via the quark-antiquark dipoles, (ii) the incident gluon 
behaves as a pair of color-uncorrelated quark and antiquark propagating at the 
same impact parameter and (iii) 
$\Phi_g(\bb,\bkappa_2) \Phi_g(\bb,\bkappa_1)$ can be considered as a
product of uncorrelated distortion factors of the quark and antiquark. 
As in the case of $q\to \{qg\}_3 $, we can treat 
$$
{1\over 2}\Big\{ |\Psi(1;z,\bp-\bkappa_1) - \Psi(z, \bp - z \bDelta)|^2 +
|\Psi(1;z,\bp-\bkappa_2) - \Psi(z, \bp - z \bDelta)|^2 
\Big\} \, ,
$$
which contains the collinear singularity of the $g \to gg$ splitting,
as an intranuclear-distorted hard fragmentation 
function of the quasielastically scattered
gluon. In close similarity to the excitation $q\to  \{qg\}_3$, one of the
wave functions which enter the fragmentation function is coherently distorted
over the whole thickness of the nucleus. The only difference is, that 
with the incident gluon, the argument 
of the coherently distorted wave function is $\bp-\bkappa_i$ versus $\bp-\bDelta$
in the case of the incident quark in $q\to  \{qg\}_3$.


\subsection{Coherent diffractive dijets and the rapidity gap survival}

The coherent diffractive back-to-back dijets form a universality class of their 
own \cite{Nonuniversality,QuarkGluonDijet}. 
They can readily be isolated from the generic octet final states upon
the application of the expansion (\ref{eq:6.1.9}):
\bea
{d \sigma_D (g^* \to \{g_1g_2 \}_{8_A} ) \over d^2\bb dz d^2\bp d^2\bDelta }  
 ={1 \over (2 \pi)^2}   \textsf{S}[\bb,\sigma_0] \delta(\bDelta)|\Psi(1;z,\bp) - \Psi(z,\bp)|^2 \, . 
\label{eq:6.2.1}
\eea
The incident gluon has a net color charge, and in close similarity 
do diffractive excitation of color-triplet quark-gluon dijets from quarks, $q\to  \{qg\}_3$,
the coherent diffractive cross section is suppressed by a nuclear attenuation factor. 
For the color-octet incident gluon the attenuation is stronger than the one 
for the color-triplet quark, $\textsf{S}[\bb,\sigma_0^{(q\bar{q})}]_{qg}$, the two
attenuation factors are related by
\bea
 \textsf{S}[\bb,\sigma_0^{(gg)}]_{gg} = \Bigl( \textsf{S}[\bb,\sigma_0^{(q\bar{q})}]_{qg}\Bigr)^{C_A/C_F}.
\label{eq:6.2.2}
\eea
Here we notice that the nuclear attenuation of coherent diffraction
can be identified with Bjorken's rapidity gap survival probability 
\cite{Bjorken}. To this end, the relationship  (\ref{eq:6.2.2}), in conjunction
with an absence of similar nuclear attenuation for coherent quark-antiquark dijets
in DIS, implies a strong breaking of diffractive factorization: the pattern
of nuclear suppression of diffraction depends strongly on the hard subprocess
and one can not treat diffractive production off nuclei as a hard interaction with
partons of a universal nuclear pomeron. A more detailed discussion of this issue and
possible implications for diffractive processes at proton-(anti)proton colliders
will be
reported elsewhere.


\subsection{Excitation of gluon-gluon dijets in symmetric octet states}

It is interesting to evaluate the contribution of symmetric octets alone.
Here the relevant excitation operator is
\bea
&& {1 \over \sqrt{2}} \Big ( \bra{e_2} - \bra{e_3} \Big) 
\textsf{S}[\bb, \hat{\Sigma}^{(0)}] \Big(  \ket{e_2} + \ket{e_3} \Big) {1 \over \sqrt{2}} 
= {1 \over 2} \Big ( \textsf{S}[\bb, \Sigma_2]- \textsf{S}[\bb, \Sigma_3] \Big ) \nonumber \\
&& =\half \textsf{S}[\bb,\T {1\over 2}\Sigma_1] \Big( \textsf{S}[\bb,  \half \sigma(\bs)] 
\textsf{S}[\bb, \half \sigma(\bs + \br - \br') ] 
\nonumber\\
&&-\textsf{S}[\bb, \T \half \sigma(\bs + \br)] \textsf{S}[\bb, \half \sigma(\bs - \br')] \Big) \, ,
\label{eq:6.3.1}\eea
which results in the cross section
\bea
&&{d \sigma (g^* \to \{g_1g_2 \}_{8_S} ) \over d^2\bb dz d^2\bp d^2\bDelta } = 
{1 \over 4 (2\pi)^2} \int d^2\bkappa_1 d^2\bkappa_2 \delta(\bDelta-\bkappa_1 -\bkappa_2)\nonumber\\
&&\times
 \Phi_g(\bb,\bkappa_1) \Phi_g(\bb,\bkappa_2) | \Psi(1;z,\bp-\bkappa_1) - \Psi(1;z,\bp-\bkappa_2) |^2 \nonumber\\
&&= {1 \over 4 (2\pi)^2} \int d^2\bkappa \phi_g(\bb,\bkappa) \phi_g(\bb,\bDelta-\bkappa) 
 | \Psi(1;z,\bp-\bkappa) - \Psi(1;z,\bp-\bDelta+\bkappa) |^2  \nonumber\\ 
&&+ {1 \over 2 (2\pi)^2}\cdot  \textsf{S}[\bb,\T {1\over 2}\sigma_0]_{gg} \phi_g(\bb,\bDelta) 
| \Psi(1;z,\bp) - \Psi(1;z,\bp-\bDelta) |^2 .
\label{eq:6.3.2}
\eea
Of course, the coherent diffractive excitation of the symmetric octet state is not allowed.


\subsection{Excitation of dijets in higher multiplets: decuplets, 27-plet, $R_7$}

In order to isolate excitation of higher multiplets we have to evaluate
the matrix element
\bea
&&- {1 \over 2} T \int_0^1  d\beta \, {N_c^2 \over \sqrt{2} N_c} \bra{e_4}  
\textsf{S}[\bb, (1- \beta)  \hat{\Sigma}^{(0)}] \,
\hat{\omega} \, \textsf{S}[\bb, \beta \hat{\Sigma}^{(0)}] \Big( \ket{e_2} +
 \ket{e_3} \Big) \nonumber \\
&& = - {N_c \over 2 \sqrt{2} } T \int_0^1  d\beta \, \bra{e_4}
\textsf{S}[\bb, (1- \beta)  \hat{\Sigma}^{(0)}] \ket{e_4} \, \bra{e_4}
\hat{\omega} \ket{e_2} \, \bra{e_2} \textsf{S}[\bb, \beta \hat{\Sigma}^{(0)}] 
\ket{e_2} \nonumber \\&& = {1 \over 4} T(\bb) \Omega(\bs,\br,\br') \, 
\int_0^1  d\beta \, \textsf{S}[\bb, (1- \beta)  \Sigma_4]  \textsf{S}[\bb, \beta \Sigma_2]  \nonumber \\
&& \equiv {1 \over 4} T(\bb) \Omega(\bs,\br,\br') \, D_A(\bb,\bs,\br,\br')
\label{eq:6.4.1}\eea
Let us concentrate on the nuclear distortion factor
\bea
D_A(\bb,\bs,\br,\br')&& =  \int_0^1  d\beta \, \textsf{S}[\bb, (1- \beta)  \Sigma_4]  \textsf{S}[\bb, \beta \Sigma_2]. 
\label{eq:6.4.2}
\eea 
It is of precisely the same form as excitation of color octet quark-antiquark 
dipoles in DIS or
excitation of sextet and 15-plet quark-gluon dipoles in quark-nucleus collisions. 
Here $\Sigma_2$ describes the initial state interactions (ISI) 
in the slice $[0,\beta]$ of the nucleus, whereas $\Sigma_4$ describes the final 
state interactions (FSI) in the slice $[\beta,1]$. Now use that fact that
$\Sigma_2={1\over 2}(\Sigma_1+\Sigma_4)$ and   
$ \textsf{S}[\bb, \beta \Sigma_2] =  \textsf{S}[\bb, {1\over 2}\beta \Sigma_4] 
\textsf{S}[\bb, {1\over 2}\beta \Sigma_1]$. Here we identify 
$\textsf{S}[\bb, {1\over 2}\beta \Sigma_1]=
\textsf{S}[\bb,  \beta \half \sigma(\br)] \textsf{S}[\bb,  \beta \half \sigma(\br')]$
with the coherent distortions of the gluon-gluon dipole wave function in the slice $[0,\beta]$,
whereas  $\textsf{S}[\bb, {1\over 2}\beta \Sigma_4]$ will give incoherent ISI 
effects in the slice  $[0,\beta]$. 

The distortion factor (\ref{eq:6.4.2}) is of precisely the same 
form as in the excitation of color octet quark-antiquark 
dipoles in DIS \cite{Nonlinear} or in the
excitation of sextet and 15-plet quark-gluon dipoles in quark-nucleus collisions
\cite{QuarkGluonDijet}. The only difference is that both the ISI and FSI cross
section operators are proportional to one and the same $\Sigma_4$, so that 
ISI and FSI can be lumped together,
\bea
 \textsf{S}[\bb, (1- \beta)  \Sigma_4] \textsf{S}[\bb, \T \half \beta  \Sigma_4] =
 \textsf{S}[\bb,  \half (2- \beta)  \Sigma_4] =
\textsf{S}[\bb,  \half \Big({C_2[27]\over C_A}(1-\beta)+\beta\Big) \Sigma_4].
\label{eq:6.4.3}\eea
Of course, in the considered large-$N_c$ approximation we have $C_2[27]=2 C_A$, 
hereafter we keep $C_2[27]$ on purpose as a reminder that collective nuclear 
glue is a density matrix in the color space, for which reason the $\beta$-dependence 
of the nontrivial effective slice $\beta+(1-\beta){C_2[27]/ C_A}$ is
controlled by the color properties of the initial and final state partons,
for a related discussion see Ref. \cite{QuarkGluonDijet}. 
The same comment is relevant to the case of decuplet dijets to be
considered in the next section.
This gives the Fourier representation 
\bea
D_A(\bb,\bs,\br,\br') & =  &  \int_0^1  d\beta \,
\textsf{S}[\bb,  \beta \T \half \sigma(\br)] 
\textsf{S}[\bb,  \beta \half \sigma(\br')] \nonumber\\
&\times&\int d^2 \bkappa_1 d^2\bkappa_2 
\exp[i \bs(\bkappa_1 + \bkappa_2)] \exp[i \bkappa_2(\br - \br')] \nonumber \\
& \times& \int_0^1  d\beta \, \Phi_g\Big({C_2[27]\over C_A}(1-\beta)+\beta,\bb,\bkappa_1\Big)  \, 
\Phi_g\Big({C_2[27]\over C_A}(1-\beta)+ \beta,\bb,\bkappa_2\Big)\nonumber\\
&= & \int_0^1  d\beta \textsf{S}[\bb,  \beta \T \half \sigma(\br)] 
\textsf{S}[\bb,  \beta \half \sigma(\br')] \D  
\int d^2 \bkappa_1 d^2\bkappa_2 d^2\bkappa_3 d^2\bkappa_4 \nonumber\\
&\times &
\exp[i \bs(\bkappa_1 + \bkappa_2 + \bkappa_3 + \bkappa_4)] \exp[i (\bkappa_2+\bkappa_4)(\br - \br')] \nonumber \\
&\times& \Phi_g\Big({C_2[27]\over C_A}(1-\beta);\bb,\bkappa_3\Big)  \Phi_g(\beta,\bb,\bkappa_1)\nonumber\\
&\times&  \Phi_g\Big({C_2[27]\over C_A}(1-\beta);\bb,\bkappa_4\Big)  \, \Phi_g(\beta,\bb,\bkappa_2). 
\label{eq:6.4.4}\eea
The second form emphasizes the distinction 
between the ISI interactions (the transverse momenta $\bkappa_{1,2}$) and FSI
(the transverse momenta $\bkappa_{3,4}$) which is obscured in the
first, convoluted, form of the distortion factor.

That gives rise to the cross section 
\bea
&&{d \sigma (g^* \to \{g_1g_2 \}_{10+ \overline{10} + 27 + R_7 } ) \over d^2\bb dz d^2\bp d^2\bDelta } = 
{1 \over 8 (2\pi)^2}\
 \int_0^1 d\beta \nonumber\\
&&\times \int d^2\bkappa_4 d^2\bkappa_3 d^2\bkappa d^2 \bkappa_1 d^2\bkappa 
\,  \delta^{(2)}(\bDelta-\bkappa- \bkappa_1 - \bkappa_2 -\bkappa_3 - \bkappa_4) \nonumber \\
&&\times f(\bkappa) \Phi_g\Big({C_2[27]\over C_A}(1-\beta);\bb,\bkappa_3\Big)  \Phi_g(\beta;\bb,\bkappa_1)
 \Phi_g\Big({C_2[27]\over C_A}(1-\beta);\bb,\bkappa_4\Big)  \, \Phi_g(\beta;\bb,\bkappa_2)\nonumber\\
&& \times 
\Big\{|\Psi(\beta;z, \bp-\bkappa_2-\bkappa_4) - \Psi(\beta;z, \bp-\bkappa_2 -\bkappa_4-\bkappa)|^2 \nonumber\\
&&+|\Psi(\beta;z, \bp-\bkappa_1-\bkappa_3) - \Psi(\beta;z, \bp-\bkappa_1 -\bkappa_3-\bkappa)|^2\Big\}
\nonumber \\
&& ={1 \over 8 (2\pi)^2} \int_0^1 d\beta 
\int  d^2\bkappa d^2 \bkappa_1 d^2\bkappa_2 f(\bkappa) 
\delta^{(2)}( \bDelta-\bkappa- \bkappa_1 - \bkappa_2) \nonumber \\
&& \times\, \Phi_g\Big({C_2[27]\over C_A}(1-\beta)+\beta;\bb,\bkappa_1\Big) \, 
\Phi_g\Big({C_2[27]\over C_A}(1-\beta)+\beta;\bb,\bkappa_2\Big) \nonumber \\
&& \times 
\Big\{|\Psi(\beta;z, \bp-\bkappa_1) - \Psi(\beta;z, \bp-\bkappa_1-\bkappa)|^2 \nonumber\\
&&+|\Psi(\beta,z, \bp-\bkappa_2) - \Psi(\beta,z, \bp-\bkappa_2-\bkappa)|^2\Big\}\,.  
\label{eq:6.4.5}
\eea
Again, we have a full agreement with the concept of the universality 
classes introduced in \cite{Nonuniversality,QuarkGluonDijet}: the pattern
of coherent distortions of the wave function and of the incoherent ISI and FSI
distortions repeats that of other processes with excitation of dijets in
color representations with the dimension higher by the factor $\propto N_c^2$ 
than the dimension of the color representation of the incident parton.


\subsection{Excitation of decuplet dijets}

The case of the decuplet dijets is exceptional because they are not
excited off free nucleons via lowest order one-gluon exchange.
A new feature is that intranuclear rescattering makes the production
of gluon dijets in the decuplets possible and it is interesting to 
look at their contribution separately.
Using $\ket{10\overline{10}} + \ket{\overline{10}10} 
= \ket{e_4}+\ket{e_5}$, the production of decuplets is 
induced by the excitation operator
\bea
&& - {1 \over 2\sqrt{2} N_c} T(\bb) \Big(\bra{e_4}+\bra{e_5}\Big) \int_0^1  d\beta \, 
\textsf{S}[\bb, (1- \beta)  \hat{\Sigma}^{(0)}] \,
\hat{\omega} \, \textsf{S}[\bb, \beta \hat{\Sigma}^{(0)}] \Big( \ket{e_2} + \ket{e_3} \Big) \nonumber \\
&& = {1 \over 4} T(\bb)\Omega(\bs,\br,\br') D^{10+\overline{10}}(\bb,\bs,\br,\br') \, ,
\label{eq:6.5.1}\eea
with the nuclear distortion factor
\bea
 D_A^{10+\overline{10}}(\bb,\bs,\br,\br') && =
\int_0^1 d\beta \Big\{ \bra{e_4}\textsf{S}[\bb, (1- \beta)  \hat{\Sigma}^{(0)}] \ket{e_4} 
\bra{e_2}  \textsf{S}[\bb, \beta \hat{\Sigma}^{(0)}] \ket{e_2} \nonumber  \\
&& -  \bra{e_5}\textsf{S}[\bb, (1- \beta)  \hat{\Sigma}^{(0)}] \ket{e_5} 
\bra{e_3}  \textsf{S}[\bb, \beta \hat{\Sigma}^{(0)}] \ket{e_3} \Big\} \nonumber\\
&&=\int_0^1 d\beta \textsf{S}[\bb, \T {1\over 2}\beta \Sigma_1] \nonumber\\
&&\times \Big\{ \textsf{S}[\bb, (1- \beta)  \Sigma_4] \textsf{S}[\bb, \T \half\beta   \Sigma_4] -
 \textsf{S}[\bb, (1- \beta)  \Sigma_5] \textsf{S}[\bb, \half\beta  \Sigma_5] \Big\}\,. 
\label{eq:6.5.2}\eea
The distinction between the ISI in the slice $[0,\beta]$ and FSI in the slice $[\beta,1]$
is obvious. Repeating the analysis in the preceding Section, we readily find 
\bea
 D_A^{10+\overline{10}}(\bb,\bs,\br,\br') && = 
\int_0^1 d\beta \, \textsf{S}[\bb, \T \half \beta  \sigma(\br)] 
\textsf{S}[\bb, \half \beta  \sigma(\br')]  \nonumber \\
&&\times \Big\{ \textsf{S}[\bb,  \T \half (2- \beta) \sigma(\bs) ]
       \textsf{S}[\bb,  \half (2- \beta)  \sigma(\bs+ \br -\br') ] \nonumber \\
&& -  \textsf{S}[\bb,  \T \half (2- \beta)  \sigma(\bs+\br) ]
\textsf{S}[\bb, \half (2- \beta)  \sigma(\bs - \br') ] \Big\} \nonumber \\
&& = \int d^2\bkappa_1 d^2\bkappa_2 \exp[i(\bkappa+\bkappa_1+\bkappa_2)\bs] \nonumber \\
&& \times \half \Big\{ \Big( \exp[i\bkappa_1\br] - \exp[i\bkappa_2\br] \Big) \, 
\Big( \exp[-i\bkappa_1 \br'] - \exp[-i \bkappa_2 \br'] \Big) \Big\} \nonumber \\
&& \times \int_0^1\! d\beta  \textsf{S}[\bb, \T  \half\beta  \sigma(\br)] 
\textsf{S}[\bb,  \half\beta \sigma(\br')] \, \nonumber \\
&& \times  \Phi_g(2-\beta,\bb,\bkappa_1)\Phi_g(2-\beta,\bb,\bkappa_2) \,.
\label{eq:6.5.3}\eea
Here
\bea
2- \beta ={C_2[10]\over C_A}(1-\beta)+\beta
\label{eq:6.5.4}\eea
and for the sake of brevity we made an explicit use of $C_2[10]=2C_A$.
The resulting contribution of the decuplet final states to the cross section
can be cast two ways
\bea
&& {d \sigma (g^* \to \{g_1g_2 \}_{10+ \overline{10}} ) \over d^2\bb dz d^2\bp d^2\bDelta }  =
 {1 \over 8 (2\pi)^2}T(\bb)
\ \int d^2\bkappa d^2 \bkappa_1 d^2\bkappa_2 \,  
\delta^{(2)}(\bkappa+ \bkappa_1 + \bkappa_2 -\bDelta) \, f(\bkappa) \nonumber \\
&& \times \int_0^1\! d\beta \, \Phi_g(2-\beta,\bb,\bkappa_1)\Phi_g(2-\beta,\bb,\bkappa_2) \nonumber \\
&& \Big\{ |\Psi(\beta,z,\bp-\bkappa_2) - \Psi(\beta,z,\bp-\bkappa_1)|^2 + 
|\Psi(\beta,z,\bp-\bkappa- \bkappa_2) - \Psi(\beta,z,\bp- \bkappa -\bkappa_1)|^2 \nonumber \\
&& + |\Psi(\beta,z,\bp-\bkappa_2) - \Psi(\beta,z,\bp-\bkappa- \bkappa_2)|^2 + 
|\Psi(\beta,z,\bp-\bkappa_1) - \Psi(\beta,z,\bp- \bkappa -\bkappa_1)|^2 \nonumber \\
&& -|\Psi(\beta,z,\bp-\bkappa_2) - \Psi(\beta,z,\bp-\bkappa- \bkappa_1)|^2 - 
|\Psi(\beta,z,\bp-\bkappa_1) - \Psi(\beta,z,\bp-\bkappa- \bkappa_2)|^2 \Big\} \nonumber\\
&&= {1 \over 8 (2\pi)^2}T(\bb)
\ \int d^2\bkappa d^2 \bkappa_1 d^2\bkappa_2 \,  
\delta^{(2)}(\bkappa+ \bkappa_1 + \bkappa_2 -\bDelta) \, f(\bkappa) \nonumber \\
&& \times \int_0^1\! d\beta \, \Phi_g(2-\beta,\bb,\bkappa_1)\Phi_g(2-\beta,\bb,\bkappa_2) \nonumber \\
&& \Big|\Psi(\beta,z,\bp-\bkappa_1) - \Psi(\beta,z,\bp-\bkappa_2) -
\Psi(\beta,z,\bp-\bkappa- \bkappa_1) + \Psi(\beta,z,\bp- \bkappa -\bkappa_2)\Big|^2 \,.
\nonumber\\
\label{eq:6.5.5}\eea
The product $ \Phi_g(2-\beta,\bb,\bkappa_1)\Phi_g(2-\beta,\bb,\bkappa_2)$
contains the term $\textsf{S}^2[\bb, \half (2-\beta)\sigma_0] \delta(\bkappa_1)\delta(\bkappa_2)$.
The similar term was the source of coherent diffraction into the antisymmetric octet.
As one would have expected, such a coherent diffractive  contribution to the decuplet cross section
vanishes.
The production of the decuplet double gluon states should lead to 
some interesting physical consequences in the sense of the final hadron state. 
In Ref. \cite{b_flow} (see also \cite{decameron,annihilation}) it was 
pointed out that the process  $g\rightarrow \{gg\}_{10}$ should
lead to the production of baryonium states and baryon number flow over a 
large rapidity gap. This observation is based on the fact that in terms of the 
triplet color spinor indices the decuplet state is described by the color 
wave function with three indices $\Psi^{ijk}$. For this reason the color 
neutralization of the decuplet state during the hadronization stage 
requires picking up from the vacuum three antiquarks (if one neglects 
purely gluon color neutralization which should be strongly 
suppressed as compared to the light quark 
mechanism due to a large effective gluon mass in the QCD vacuum 
$m_{g}^{eff}\sim R_{c}^{-1}$, where $R_{c}\sim 0.27$ fm is the gluon 
correlation radius in the vacuum \cite{Shuryak}). In the string model 
\cite{strings} the hadronization proceeds through the breaking of the triplet 
color strings (three string for the decuplet state) due to Schwinger 
production of $q\bar{q}$ pairs in the color-electric field of the triplet 
strings \cite{Nussinov}\footnote{In the string model for $N_{c}=3$ baryon is 
usually described by the Y-configuration of three triplet strings connected in
the so-called string junction, which plays the role of a carrier of the baryon 
number \cite{Veneziano}. The baryonium state is a system of 
junction-antijunction connected by three triplet color strings.}.
For the decuplet double gluon state the baryon 
number $N_{B}=-1$ will be compensated by production of an additional baryon
in the nucleus fragmentation region. In the case of the antidecuplet state 
we have $N_{B}=1$ in the double gluon rapidity region, which can be viewed as
a flow of valence baryon number from the nucleus region to the double gluon region.
This effect may be important for the baryon stopping in $pA$ and 
$AA$ collisions. The corresponding numerical estimates will be given elsewhere.
In Ref. \cite{z_moriond98} it was pointed out that a similar mechanism 
with gluon splitting into color decuplet/antidecuplet 
double gluon states in the quark-gluon plasma produced in the initial stage of 
$AA$ collisions should increase the high-$p_{T}$ baryon production
in ultrarelativistic heavy ion collisions. Numerical calculations 
\cite{Aurenche} show that this mechanism (and the processes 
$q\rightarrow \{qg\}_{\bar{6}}$ and 
$\bar{q}\rightarrow \{\bar{q}g\}_{6}$, discussed in \cite{QuarkGluonDijet}, which 
also lead to baryon production) may really play    
an important role in the anomalously large 
baryon/meson ratio observed experimentally at RHIC (for the recent review
see \cite{STARreview}).
 

\section{Conclusions}

We derived the nonlinear 
$k_\perp$-factorization for the last missing pQCD subprocess -
production of hard gluon-gluon dijets in gluon-nucleus collisions
when the nuclear coherency condition $x\lsim x_A \approx 
0.1\cdot A^{-1/3}$ holds. Although of limited importance 
at not so high energies of RHIC, this subprocess
will be a dominant source of mid-rapidity and proton-hemisphere
dijets in $pA$ collisions at LHC. The principal technical novelty is a
solution of the rather involved seven-channel
non-Abelian evolution equations for intranuclear propagation
of four-gluon states.
Our results for the  gluon-gluon dijets in all color representations 
are presented in the form of explicit quadratures.
The concept of universality classes
\cite{Nonuniversality,QuarkGluonDijet} is fully corroborated:
The nonlinear $k_\perp$ factorization properties of excitation of 
digluons in higher color representations are identical to 
those in excitation of color-octet quark-antiquark dijets in
DIS and quark-gluon dijets in higher color
representations in $qA$ collisions. Similar nonlinear 
$k_\perp$ factorization properties are exhibited by 
excitation of dijets in the same color 
representation as the incident parton: $g\to  \{gg\}_8,\, g\to \{q\bar{q}\}_8,
\, q\to  \{qg\}_3$. In both $g\to gg$ and $q\to qg$ processes 
coherent diffractive excitation of incident partons
with net color charge is suppressed by a nuclear absorption 
factor which can be identified with Bjorken's gap survival probability. 
The gap survival probabilities in the two cases are different.
Furthermore, the related absorption is
absent in diffractive $\gamma^*\to q\bar{q}$, which is
indicative of a strong breaking of diffractive factorization.
We mentioned mid-rapidity to proton hemisphere gluon-dijets dijets
in $pA$ collisions at LHC as a future application of the derived
formalism. Still another potential application of our results
for color-decuplet digluon production is a baryon number flow
from the nucleus to large rapidity region. But, first and foremost,
this work completes a derivation of nonlinear $k_\perp$-factorization
for all pQCD processes in a nuclear medium  and opens a way to a systematic
compartive studies of high-$p_\perp$ jet-jet and hadron-hadron
correlations in different parts of the phase space of DIS off
nuclei and hadron-nucleus collisions.

The diagonalization properties of the single-jet
problem are somewhat beyond the major theme of this communication. 
Still, in view of the discussion in sec. VI.E the isolation of 
contributions from different final states 
is of certain interest, and we included Appendix C. The
manifest diagonalization of the initially seven-coupled channel 
problem in the $t$-channel basis is, apparently, of more general 
interest and may find further applications in other problems.

{\bf Acknowledgements}: This work was partly supported by  grants 
DFG 436 RUS 17/101/04 and DFG RUS 17/138/05.
\newpage

\section*{Appendix A: Useful $SU(N_c)$ relations}
\setcounter{equation}{0}
\renewcommand{\theequation}{A.\arabic{equation}}

In this appendix 
we collect a number of
identities that are helpful in the evaluation of the matrix 
elements of the dipole cross section operator.
In the derivation of the projectors we  follow closely, 
though in slightly different notation and normalization,
Ref. \cite{Predrag}. Many useful $SU(N_c)$--identities can be found in
Refs. \cite{MacFarlane:1968vc,ColorAlgebra}.

If $t^a, a = 1 \dots N_c^2-1$
are $SU(N_c)$-generators in the fundamental representation, 
the familiar $f$-- and $d$--tensors are defined through
\bea
t^a t^b = {1 \over 2 N_c} \delta_{ab} \, \openone 
+ { 1\over 2} \Big( d_{abc} + i f_{abc} \Big) t^c \, ,
\label{eq:A.1}
\eea
or, 
\bea
i f_{abc} = 2 \, \Tr\Big([t^a,t^b]t^c\Big) \, , \, d_{abc} = 2 \, \Tr\Big(\{t^a,t^b\}t^c\Big) \, .
\label{eq:A.2}
\eea
The $SU(N_c)$--generators in the adjoint representation are
\bea
(T^a)_{bc} = i f_{bac} \, 
\label{eq:A.3}
\eea
so that their defining property $[T^a,T^b] = i f_{abc} T^c$,
and the $SU(N_c)$ transformation properties of $d$--symbols
give rise to the Jacobi identities
\bea
if_{kam}if_{mbl} -if_{kbm}if_{mal} = i f_{abm} if_{kml} \, ,
\label{eq:A.4}
\eea
\bea
f_{kam} d_{mbl} - d_{kbm} f_{mal} =  f_{abm} d_{kml} \, .
\label{eq:A.5}
\eea

To evaluate contractions of multiple $f$-- and $d$--symbols one makes
use of the Fierz--identity for the fundamental generators,
\bea
(t^a)^i_j (t^a)^k_l = {1 \over 2} \delta^i_l \delta^k_j - {1 \over 2N_c} \delta^i_j \delta^k_l \, .
\label{eq:A.6}
\eea
They entail, that 
\bea
\Tr\Big(At^a\Big) \Tr\Big(Bt^a\Big) = {1 \over 2} \Tr( AB ) - {1 \over 2N_c} \Tr(A) \Tr(B) \nonumber \\
\Tr\Big(At^aBt^a\Big) = {1 \over 2} \Tr(A) \Tr(B)  -  {1 \over 2N_c}\Tr( AB ) \, .
\label{eq:A.7}
\eea
In conjunction with eq.(\ref{eq:A.2}) one can then obtain

\bea
f_{aij}f_{bij} &&= N_c \, \delta_{ab} \, , \nonumber \\ 
 d_{aij}d_{bij} &&= {N_c^2 - 4 \over N_c} \delta_{ab} \, , \\ 
 f_{aij}d_{bij} &&= 0 \, . \nonumber
\label{eq:A.8}
\eea
\bea
f_{iaj}f_{jbk}f_{kci} &&= -{N_c \over 2} f_{abc} \, , \nonumber \\
f_{iaj}f_{jbk}d_{kci} &&= -{N_c \over 2} d_{abc} \nonumber \\
f_{iaj}d_{jbk}d_{kci} &&= {N_c^2 - 4  \over 2N_c} f_{abc} \\  
d_{iaj}d_{jbk}d_{kci} &&= {N_c^2 - 12 \over 2N_c} d_{abc} \, . \nonumber
\label{eq:A.9}
\eea
\bea
f_{kan}f_{nbm}f_{mdl}f_{lck} &&= \delta_{ac}\delta_{bd} + \delta_{ab} \delta_{cd} 
+ {N_c \over 4} ( d_{ack}d_{kdb} + d_{abk}d_{kcd} - d_{adk}d_{kbc} ) \nonumber \\
d_{kan}d_{nbm}d_{mdl}d_{lck} &&= {N_c^2 -4 \over N_c^2} 
\Big( \delta_{ac}\delta_{bd} + \delta_{ab} \delta_{cd} \Big) + 
{N_c^2 - 16 \over 4N_c} \Big(d_{ack}d_{kdb} + d_{abk}d_{kcd} \Big) \nonumber \\
&&- {N_c\over 4} d_{adk}d_{kcb} \, . 
\label{eq:A.10} 
\eea

It is helpful to analyse the box and twisted-box traces of
four fundamental generators,
\bea
R^{ab}_{cd} \equiv 4 \, \Tr\Big(t^a t^b t^d t^c \Big) \, , \, Q^{ab}_{cd} = 
4 \, \Tr\Big(t^a t^d t^b t^c \Big) \, .
\label{eq:A.11}\eea
From eq.(\ref{eq:A.1}), we obtain immediately
\bea
R_{cd}^{ab} = {1 \over N_c} \delta_{ab}\delta_{cd} + 
{1 \over 2} ( d_{abk}d_{kdc} + i f_{abk} i f_{kdc} )+ 
{i \over 2} ( d_{abk}f_{kdc} + f_{abk} d_{kdc} ) \, ,
\label{eq:A.12}
\eea
on the other hand,
\bea
R_{cd}^{ab} \equiv 4 \, \Tr\Big(t^a t^b t^d t^c \Big) =  4 \, \Tr\Big(t^c t^a t^b t^d \Big) \, ,
\label{eq:A.13}
\eea
so that also
\bea
R_{cd}^{ab} = {1 \over N_c} \delta_{ac}\delta_{bd} + 
{1 \over 2} ( d_{cak}d_{kbd} + i f_{cak} i f_{kbd}) + 
{i \over 2} ( d_{cak}f_{kbd} + f_{cak} d_{kbd} ) \, .
\label{eq:A.14}
\eea
We can equate the real and imaginary parts of $R$ from eqs.(\ref{eq:A.12},\ref{eq:A.14}) 
separately, and thus obtain the identities
\bea
  d_{abk}f_{kdc} + f_{abk} d_{kdc} =  d_{cak}f_{kbd} + f_{cak} d_{kbd} \, ,
\label{eq:A.15}
\eea
\bea
{2 \over N_c} (\delta_{ac}\delta_{bd} -  \delta_{ab}\delta_{cd}) +  d_{cak}d_{kbd} - d_{abk}d_{kdc}
 &&=  i f_{abk} i f_{kdc} - i f_{cak} i f_{kbd} \nonumber \\
 &&= i f_{adk} i f_{kbc}  \, .
\label{eq:A.16}
\eea
For the tensor $Q$ we have 
\bea
Q_{cd}^{ab} &&= {1 \over N_c}  \delta_{ad}\delta_{bc} + 
{1 \over 2} ( d_{adk}d_{kbc} + i f_{adk} i f_{kbc}) + 
{i \over 2} ( d_{adk}f_{kbc} + f_{adk} d_{kbc} ) \nonumber \\
&&= {1 \over N_c} ( \delta_{ad}\delta_{bc} + \delta_{ac}\delta_{bd} -  \delta_{ab}\delta_{cd} ) +
{1 \over 2} ( d_{adk}d_{kbc} +  d_{cak}d_{kbd} - d_{abk}d_{kdc}) + iY^{ab}_{cd} \, , 
\label{eq:A.17}
\eea
where we introduced a shorthand notation
\bea
 iY^{ab}_{cd} &&=  {i \over 2} \Big( d_{adk}f_{kbc} + f_{adk} d_{kbc} \Big) \nonumber \\ 
&&=  {i \over 2} \Big( f_{cak} d_{kdb}+  d_{cak}f_{kdb} \Big) \, ,
\label{eq:A.18}
\eea
and made use of eqns.(\ref{eq:A.15},\ref{eq:A.16}).

\section*{Appendix B: Derivation of the projectors onto irreducible representations}
\setcounter{equation}{0}
\renewcommand{\theequation}{B.\arabic{equation}} 

Our task is now to find the irreducible representations  (\ref{eq:4.0.1}) for the product 
of two adjoints and to construct the relevant projection operators.
The auxiliary tensors: 
\bea
{\cal{S}}^{ab}_{cd}  \equiv {1 \over 2}\Big( \delta_{ac} \delta_{bd} + \delta_{ad} \delta_{bc} \Big)\,, \, 
{\cal{A}}^{ab}_{cd}  \equiv {1 \over 2}\Big( \delta_{ac} \delta_{bd} - \delta_{ad} \delta_{bc} \Big) \, 
\label{eq:B.1}
\eea
 decompose the product representation space into its 
symmetric and antisymmetric parts, respectively:
\bea
 \openone^{ab}_{cd} \equiv  \delta_{ac} \delta_{bd} = {\cal{S}}^{ab}_{cd} + {\cal{A}}^{ab}_{cd}
\label{eq:B.2}
\eea
In addition, also 
\bea
[D_t]^{ab}_{cd} \equiv d_{ack}d_{kbd}\, , \, [D_u]^{ab}_{cd} \equiv d_{adk}d_{kbc} \, ,
[D_s]^{ab}_{cd} \equiv d_{abk}d_{kcd}
\label{eq:B.3}\eea
will prove helpful. All the above defined tensors ${\cal{S}},{\cal{A}}, D_s,D_t,D_u$,
as well as $iY$ of eq.(\ref{eq:A.18}) 
are hermitian, i.e. 
$({\cal{O}}^\dagger)^{ab}_{cd} =  ({\cal{O}}^{cd}_{ab})^*={\cal{O}}^{ab}_{cd} $.
The $SU(N_c)$-projectors onto the singlet as well as the 
two adjoint multiplets have up to the normalization factors 
manifestly the same form as their well-known $N_c=3$ counterparts.
\bea
P[1]^{ab}_{cd} &=& {1 \over N_c^2-1} \delta_{ab} \delta_{cd} \\
P[8_S]^{ab}_{cd} &=& {N_c \over N_c^2-4}  d_{abk}d_{kcd} =  {N_c \over N_c^2-4}\, [D_s]^{ab}_{cd}
\label{eq:B.4}\eea
project onto the symmetric singlet and octet, respectively, while
\bea
P[8_A]^{ab}_{cd} &=& {1 \over N_c} f_{abk}f_{kcd} = {1 \over N_c} i f_{abk} i f_{kdc}
\label{eq:B.5}\eea
projects onto the antisymmetric octet.
It is easily checked that they are indeed hermitian and satisfy
the requirement
\bea
(P[R_i]^2)^{ab}_{cd} = P[R_i]^{ab}_{kl} P[R_i]^{kl}_{cd} = P[R_i]^{ab}_{cd}  
\label{eq:B.6}\eea
Using the identities (\ref{eq:A.8},\ref{eq:A.9}) 
one finds the number of states they propagate:
\bea
\Tr P[R_i] \equiv  P[R_i]^{ab}_{ab} = \dim[R_i] \, ,
\label{eq:B.7}\eea
explicitly,
\bea 
\Tr P[1] = 1, \, \,   \Tr P[8_A] = \Tr P[8_S] = N_c^2-1 \, ,
\label{eq:B.8}\eea
The symmetric and antisymmetric parts of our space
contain, respectively
\bea
\Tr {\cal{S}} = {1 \over 2} N_c^2 (N_c^2-1) \, , \Tr {\cal{A}} = {1 \over 2} (N_c^2-1)(N_c^2-2) \, 
\label{eq:B.9}\eea
states, so that now the problem arises to find the decomposition into
irreducible representations of the subspaces that belong to the projectors
\bea
{\cal{S}}_\perp = {\cal{S}}- P[1] - P[8_S] \, \, , \, \, {\cal{A}}_\perp = {\cal{A}}- P[8_A] \, .
\label{eq:B.10}\eea
This is done most straightforwardly, following \cite{Predrag}, by 
investigating the above defined tensor $Q$ (\ref{eq:A.17}), which takes the form
\bea
Q = {2 \over N_c} {\cal{S}} - {N_c^2-1 \over N_c} P[1] - {N_c^2-4 \over 2N_c} P[8_S] + 
{1 \over 2} ( D_u + D_t) + i Y
\label{eq:B.11}\eea
First notice that its symmetric (antisymmetric) part is purely real (imaginary),
\bea
{\cal{S}}Q{\cal{S}} = \Re e \, Q \, ,\, {\cal{A}}Q{\cal{A}} = i \Im m \, Q \, ,
\label{eq:B.12}\eea
and furthermore
\bea
{\cal{S}}Q{\cal{S}} = {\cal{S}}Q = Q{\cal{S}} \, , \, 
{\cal{A}}Q{\cal{A}} = {\cal{A}}Q = Q{\cal{A}} \, .
\label{eq:B.13}\eea
Now evaluate its square, $Q^2$ -- best by starting from its definition
as a trace of fundamental generators --
\bea
(Q^2)^{ab}_{cd} &&= 16 \, \Tr(t^a t^l t^b t^k)\Tr(t^k t^d t^l t^c) \nonumber \\
&&= 4 \Tr(t^a t^c) \Tr (t^b t^d) + {4 \over N_c^2} \Tr(t^ct^d)\Tr(t^at^b) - {4 \over N_c}
\Big( \Tr (t^at^bt^dt^c) + \Tr(t^c t^d t^bt^a) \Big) \nonumber \\
&&= \delta_{ac} \delta_{bd} + {1 \over N_c^2} \delta_{cd}\delta_{ab} - {2 \over N_c} 
\Re e \, R_{ab}^{cd} \, ,
\label{eq:B.14}\eea
so that 
\bea
Q^2 = \openone - {N_c^2-1 \over N_c^2} P[1] - {N_c^2-4 \over N_c^2} P[8_S] - P[8_A]  \, .
\label{eq:B.15}\eea
From here we can conclude, that on the subspaces under investigation,
$Q^2$ acts as the unit matrix:
\bea
{\cal{S}}_\perp Q^2 =   Q^2 {\cal{S}}_\perp = {\cal{S}}_\perp \, , \, 
{\cal{A}}_\perp Q^2 =   Q^2 {\cal{A}}_\perp = {\cal{A}}_\perp \, ,
\label{eq:B.16}\eea
therefore both subspaces decompose into orthogonal eigen-spaces
belonging to eigenvalues $\pm 1$ of the operator $Q$.
We can then write down the projection operators
\bea
P_{A_\perp}^{\pm} &&= {1 \over 2} \, (\openone \pm Q )  {\cal{A}}_\perp 
= {1 \over 2} \, (\openone \pm i \,  \Im m\,  Q )  {\cal{A}}_\perp  
\label{eq:B.17}\eea
\bea
P_{S_\perp}^{\pm} && = {1 \over 2} \, (\openone \pm Q )  {\cal{S}}_\perp =  
{1 \over 2} \, (\openone \pm  \,  \Re e\,  Q )  {\cal{S}}_\perp \, . 
\label{eq:B.18}\eea

To evaluate the dimensions of the associated representations, we
first derive more explicit forms of the projectors.
We start with the symmetric case. From the relations 
(\ref{eq:A.8},\ref{eq:A.9}) we obtain
\label{eq:B.19}\bea
(D_u + D_t) P[1] = {2 (N_c^2-4) \over N_c} P[1] \, , \, (D_u + D_t) P[8_S] = {N_c^2-12 \over N_c} P[8_S] \, ,
\label{eq:B.20}\eea
and, trivially,
\bea
 (D_u + D_t) {\cal{S}} =  D_u + D_t \, ,
\label{eq:B.21}\eea
so that
\bea
P_{S_\perp}^{\pm} &&= {1 \over 2} \Big\{ \Big(1 \pm{2 \over N_c} \Big) 
{\cal{S}}_\perp \pm {1\over 2}
\Big( D_u + D_t \Big) \mp {N_c^2 -4 \over N_c} P[1] 
\mp {N_c^2-12 \over 2N_c} P[8_S] \Big\} \nonumber \\
&&= {1 \over 2} \Big\{ \Big(1 \pm {2\over N_c} \Big) 
{\cal{S}} \mp {(N_c \pm2)(N_c \mp1) \over N_c} P[1] \mp 
{(N_c \mp2)(N_c \pm4) \over 2N_c} P[8_S] \pm {1\over2}\Big(D_u + D_t\Big) \Big\} \, . 
\nonumber \\
\label{eq:B.22}\eea
Now,
\bea
\Tr D_t = 0 \, , \, \Tr D_u = {(N_c^2 -4)(N_c^2-1) \over N_c} \, ,
\label{eq:B.23}\eea
so that we can easily establish, that
\bea
\Tr P_{S_\perp}^{+} = {N_c^2 (N_c-1) (N_c+3) \over 4}  \, , \, 
\Tr P_{S_\perp}^{-} = {N_c^2 (N_c+1) (N_c-3) \over 4}
\label{eq:B.24}\eea
For $N=3$, we have
\bea
\Tr P_{S_\perp}^{+}\Big|_{N_c=3} = 27  \, , \, 
\Tr P_{S_\perp}^{-}\Big|_{N_c=3} = 0 \, ,
\label{eq:B.25}\eea
so that from now on we shall denote $P_{S_\perp}^{+} \equiv P[27]$, whereas
for the other symmetric representation, which vanishes for $N_c=3$ we shall use
the notation $P_{S_\perp}^{-} \equiv P[R_7]$.
It is interesting to note, that the vanishing of $R_7$ for three colors 
can be related to a well--known accidental cancellation, namely 
\bea
P[R_7]^{ab}_{cd}\Big|_{N_c=3} = {1 \over 4} \Big\{ {1 \over 3} 
\Big(  \delta_{ac} \delta_{bd} + \delta_{ad} \delta_{bc} +  
\delta_{ab} \delta_{cd} \Big) - \Big( d_{ack}d_{kbd} + 
d_{adk}d_{kbc} + d_{abk}d_{kcd} \Big) \Big\} \, , \nonumber \\
\label{eq:B.26}\eea
is identically zero for $SU(3)$ \cite{MacFarlane:1968vc}. 

This completes the reduction of the symmetric part, where we have 
\bea
{\cal{S}} = P[1] + P[8_S] + P[27] + P[R_7] \, .
\label{eq:B.27}\eea 
We now turn to the antisymmetric part of the product representation space,
where we deal with two complex conjugate multiplets.
Here we see, that
\bea
Q P[8_A] = i \Im m  Q P[8_A] = iY P[8_A] = 0 \, ,
\label{eq:B.28}\eea
and hence
\bea
P_{A_\perp}^{\pm} &&= {1 \over 2} \, (\openone \pm Q )  {\cal{A}}_\perp 
= {1 \over 2} \, (\openone \pm i \,  \Im m \,  Q )  {\cal{A}}_\perp \nonumber \\
&&=  {1 \over 2} \, \Big( {\cal{A}} - P[8_A] \pm i  Y \Big) \, .
\label{eq:B.29}\eea
As $\Tr \, iY = 0 $, we have
\bea
\Tr P_{A_\perp}^{\pm} = {(N_c^2 - 1)(N_c^2 -4) \over 4} \, , 
\label{eq:B.30}\eea
for $SU(3)$
\bea
\Tr P_{A_\perp}^{\pm}\Big|_{N_c=3} = 10 \, ,
\label{eq:B.31}\eea
so that from now on
\bea
P_{A_\perp}^{+} \equiv P[10] \, , \, P_{A_\perp}^{-} \equiv P[\overline{10}] \, .
\label{eq:B.32}\eea
Of course in this context it is merely a convention which multiplet we address as
the decuplet and which as the antidecuplet.
This completes our reduction of the antisymmetric part,
\bea
{\cal{A}} = P[8_A] +   P[10] +  P[\overline{10}] \, .
\label{eq:B.33}\eea

\section*{Appendix C: Eigenstates for the single-particle problem and the crossing matrix}
\setcounter{equation}{0}
\renewcommand{\theequation}{C.\arabic{equation}}

\begin{figure}[!t]
\begin{center}
\includegraphics[width = 4 cm,height= 6 cm, angle=270]{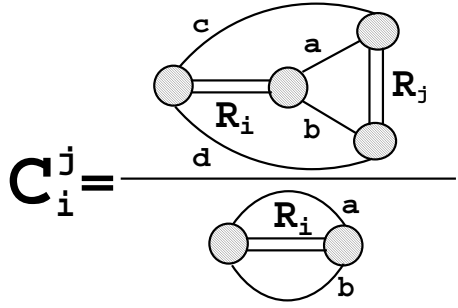}
\caption{The matrix element $C^j_i$ of the crossing matrix.}
\label{fig:Crossing_diagram}
\end{center}
\end{figure}
In sec. IV we computed the dipole cross section operator 
in the $s$-channel basis of states where gluons $1,2$ and $1',2'$ 
respectively were in a definite
$SU(N_c)$--multiplet. For the remainder of this section let us denote 
these states by $\ket{R\overline{R}}_s$:
\bea
\ket{R\overline{R}}_s && \equiv \ket{\Big\{ [g^a(\bb_1) \otimes g^b(\bb_2) \Big]_R 
\otimes [g^c(\bb'_1) \otimes g^d(\bb'_2) \Big]_{\overline{R}} \Big\}_1 } 
\nonumber \\ 
&&=  {1 \over \sqrt{\dim
[R]}} \,P[R]^{ab}_{cd} \, \ket{g^a(\bb_1) \otimes g^b(\bb_2) 
\otimes g^c(\bb'_1) \otimes g^d(\bb'_2)} \, .
\label{eq:C.1}\eea
We mentioned, that for purposes of the single particle spectrum and total
cross section, the dipole cross section operator is diagonalized 
in the $t$-channle basis where gluons $1,1'$ and $2,2'$ respectively are in definite
color multiplets, we shall denote this basis by
$\ket{R\overline{R}}_t$:
\bea
\ket{R\overline{R}}_t && \equiv \ket{\Big\{ [g^a(\bb_1) \otimes g^c(\bb'_1) \Big]_R 
\otimes [g^b(\bb_2) \otimes g^d(\bb'_2) \Big]_{\overline{R}} \Big\}_1 } 
\nonumber \\ 
&&=  {1 \over \sqrt{\dim
[R]}} \,P[R]^{ac}_{bd} \, \ket{g^a(\bb_1) \otimes g^b(\bb_2) 
\otimes g^c(\bb'_1) \otimes g^d(\bb'_2)} \, .
\label{eq:C.2}
\eea

The proof, that the basis (\ref{eq:C.2}) indeed diagonalizes the 
dipole cross section operator of the single particle spectrum
proceeds as follows. First, the dipole cross section matrix w.r.t.
the basis (\ref{eq:C.2}) is obtained from the one w.r.t. 
(\ref{eq:C.1}) by the swap of impact parameters
\bea
\{\bb_1,\bb_2,\bb'_1, \bb'_2 \}  \to \{\bb_1,\bb'_1,\bb_2, \bb'_2\} \, .
\label{eq:C.3}
\eea
That means, for the off--diagonal piece we obtain
\bea
_t\bra{R'\overline{R'}} \hat{\sigma}^{(4)} \ket{R \overline{R}}_t \propto 
\Omega (\bb_1,\bb'_1,\bb_2, \bb'_2 ) = \sigma(\bb'_1 - \bb_2) + \sigma(\bb_1 - \bb'_2)
- \sigma(\bb_1 - \bb_2) - \sigma(\bb'_1 - \bb'_2) \, \, . \nonumber \\
\label{eq:C.4}\eea
Now, for the single particle spectrum, one would integrate out,
say $\bp_2$ in the master formula (\ref{eq:2.2.8}), and in effect put $ \bb_2 = \bb'_2$,
but then, 
\bea
\Omega (\bb_1,\bb'_1,\bb_2, \bb_2 ) \equiv 0 \, . 
\label{eq:C.5}\eea
Hence, for the purposes of the single particle spectrum,
the off--diagonal elements of the dipole cross section 
vanish identically in the basis of states $\ket{R \overline{R}}_t $,
which is what we set out to prove.

As the dipole cross section matrix in the basis $(\ref{eq:C.2})$ is
obtained from the simple swap (\ref{eq:C.3}), we can immediately give its 
eigenvalues $\lambda_i$:
\bea
\lambda_1 &&= \sigma(\br-\br') \nonumber \\
\lambda_2 &&= \lambda_3 = {1 \over 2} \Big[ \sigma(\br) + \sigma(\br') + \sigma(\br-\br') \Big] 
\nonumber \\
\lambda_4 &&= \lambda_5 =  \sigma(\br) + \sigma(\br') \nonumber \\
\lambda_6 &&= {N_c+1 \over N_c} \Big[\sigma(\br) + \sigma(\br')\Big] - 
{ 1\over N_c} \sigma(\br - \br') \nonumber \\
\lambda_7 &&=  {N_c-1 \over N_c} \Big[\sigma(\br) + \sigma(\br')\Big] + { 1\over N_c} \sigma(\br - \br') \, ,
\label{eq:C.6}\eea 
where, as throughout the main body of the text, $\br = \bb_1 - \bb_2 \, , \, 
\br' = \bb'_1 - \bb'_2 $.
The system of eigenvectors which belong to the $\lambda_i $ is
\bea
&&\ket{\lambda_1} = \ket{11}_t \, , \, \ket{\lambda_2} = \ket{8_A 8_A}_t \, , \, 
\ket{\lambda_3} = \ket{8_S 8_S}_t \nonumber \\
&&\ket{\lambda_4} =   \ket{10 \overline{10}}_t  \, , \, 
\ket{\lambda_5} =   \ket{\overline{10} 10}_t \, , \, \ket{\lambda_6} = \ket{2727}_t \, , \,
 \ket{\lambda_7} = \ket{R_7R_7}_t 
\label{eq:C.7}\eea 

Clearly, once the spectrum of a matrix is known, the Sylvester formula
would allow one to calculate any function of the matrix without knowledge
of the eigenstates. In practice however explicit knowledge of the latter is
helpful. To obtain the color-wavefunctions of the states (\ref{eq:C.2}),
we need to establish the Fierz--type identities:
 \bea
 P_t[R_j] = \sum_{i=1}^9 C^j_i \, P_s[R_i] \, ,
\label{eq:C.8}\eea
i.e. 
\bea
 P[R_j]^{ac}_{bd}    = \sum_{i=1}^9 C^j_i \, P[R_i]^{ab}_{cd} \, ,
\label{eq:C.9}\eea
The $t$--channel projectors thus read, component--wise: 
\bea
 P_t[R]^{ab}_{cd} \equiv  P_s[R]^{ac}_{bd} \, .
\eea

The crossing matrix $C^j_i$ is now
obtained as (for a diagrammatic representation, see Fig.(\ref{fig:Crossing_diagram}))
\bea
 C^j_i =  {P[R_j]^{ac}_{bd}\cdot P[R_i]^{cd}_{ab} \over  P[R_i]^{ab}_{cd}P[R_i]^{cd}_{ab}} 
= {P[R_j]^{ac}_{bd}\cdot P[R_i]^{cd}_{ab} \over \dim[R_i]} \, . 
\label{eq:C.10}\eea

Apart from the complex, but hermitian structure 
$i Y_s = P_s[10] - P_s[\overline{10}],$ explicitly 
\bea
i (Y_s)^{ab}_{cd} = {i \over 2} \Big( f_{cak}d_{kdb} + d_{cak}f_{kdb} \Big) \, , 
\label{eq:C.11}\eea
which already appeared in the decuplet projectors, the full set of color-singlet 
four-gluon states includes two more complex,  but hermitian, tensor structures
\bea
&&i (Z_s^{(+)})^{ab}_{cd} =  {i \over 2} \Big( f_{bak} d_{kcd} + d_{bak}f_{kcd} \Big) \, , \nonumber \\
&&i (Z_s^{(-)})^{ab}_{cd} =  {i \over 2} \Big( f_{bak} d_{kcd} - d_{bak}f_{kcd} \Big) \, .
\label{eq:C.12}
\eea
These new tensors $i Z_s^{(\pm)}$ correspond to mixed $\ket{8_A 8_S}$--states.
We explicitly introduce the normalised states
\bea
&&\ket{(8_A 8_S)^+}_s = \sqrt{ 2 \over (N_c^2 -4)(N_c^2 -1)} i (Z_s^{(+)})^{ab}_{cd} \ket{g^a(\bb_1) \otimes g^b(\bb_2) 
\otimes g^c(\bb'_1) \otimes g^d(\bb'_2)}\, ,\nonumber\\
\label{eq:C.13}
&&\ket{(8_A 8_S)^-}_s = \sqrt{ 2 \over (N_c^2 -4)(N_c^2 -1)} i (Z_s^{(-)})^{ab}_{cd} \ket{g^a(\bb_1) \otimes g^b(\bb_2) 
\otimes g^c(\bb'_1) \otimes g^d(\bb'_2)} \, . \nonumber \\
\label{eq:C.14}\eea
As mentioned in the main text, these states decouple in the dipole 
cross section operator (like $iY$) from
the states $\ket{R\overline{R}}_s$ relevant to our problem.
It is straightforward to establish their crossing properties, namely
\bea
i Y_t = i Z^{(-)}_s \, , \, i  Z^{(-)}_t = i Y_s \, , \,  i  Z^{(+)}_t = - i  Z^{(+)}_s \, .
\label{eq:C.15}\eea
 
Using the projectors derived in appendix B we then obtain for the crossing matrix, 
including
the complex tensors (\ref{eq:C.12}), the result shown in eq. (C.16), and from
there also  the basis of eigenstates $\ket{R \overline{R}}_t$ 
displayed in eq (C.17). An $SU(3)$ counterpart of the crossing matrix (C.16) 
can be found, e.g. in \cite{CrossingMatrix}. Apparently, the crossing matrix could
be used for an alternative derivation of the four--gluon dipole cross section matrix.

\pagebreak

\begin{figure}[!h]
\begin{center}
\includegraphics[width = 15 cm, height = 22 cm, angle=180]{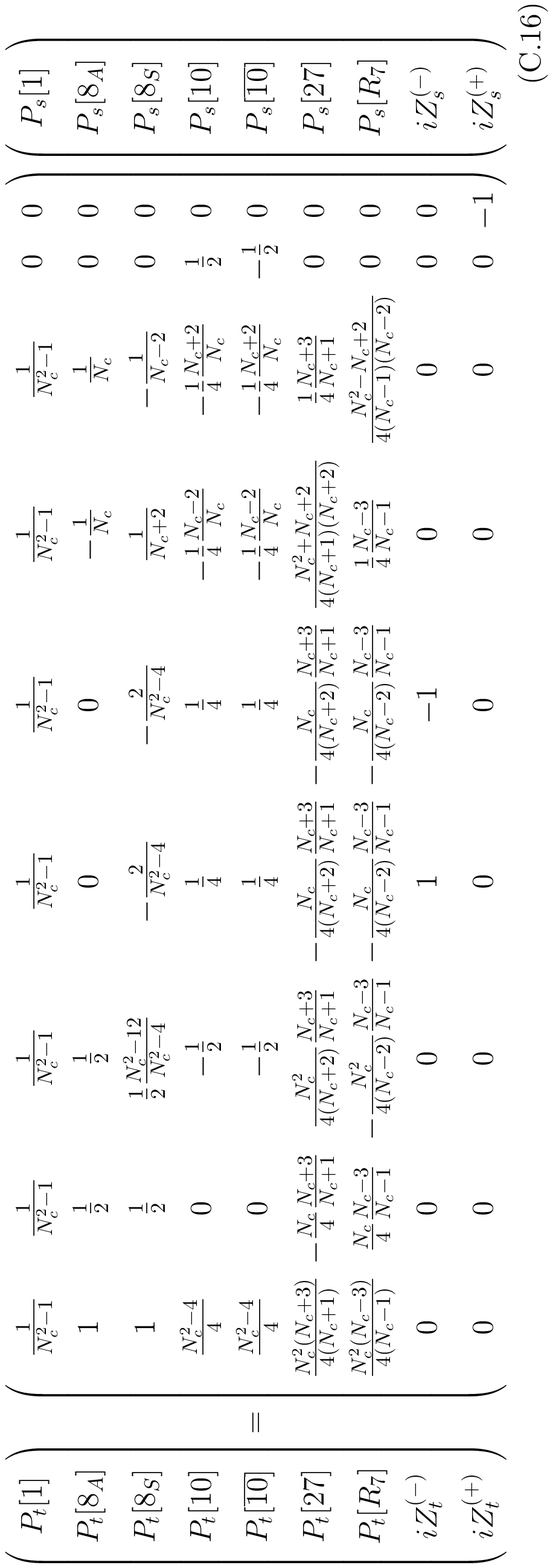}
\label{fig:CrossingMatrix}
\end{center}
\end{figure}

\pagebreak

\begin{figure}[!t]
\begin{center}
\includegraphics[width = 18 cm,height= 20 cm, angle= 180]{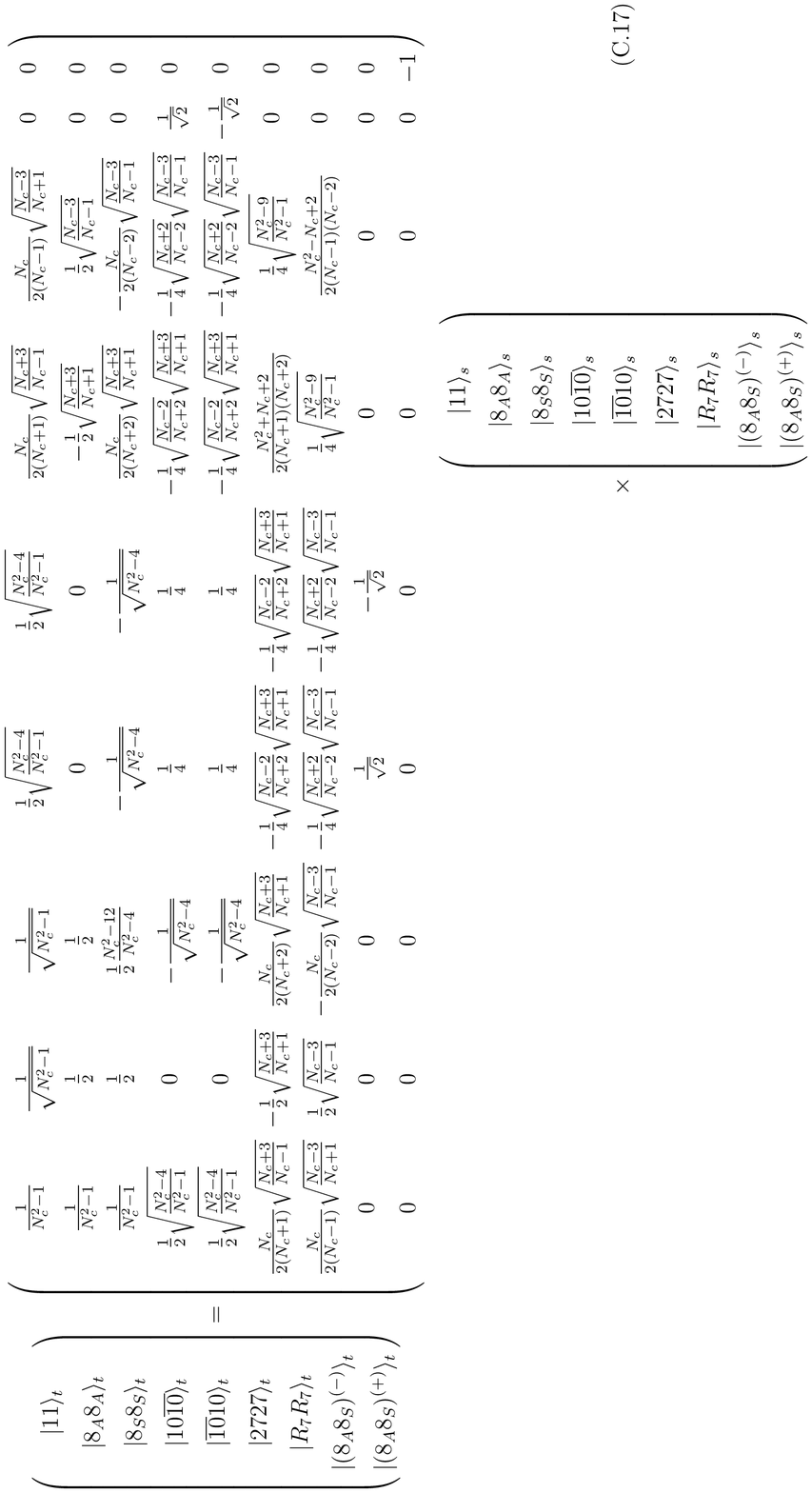}
\label{fig:Eigenstates}
\end{center}
\end{figure}

\pagebreak

\section*{Appendix D: The menagerie of nuclear collective unintegrated glue}
\setcounter{equation}{0}
\renewcommand{\theequation}{D.\arabic{equation}} 

A pertinent quantity which emerges in the description of
hard processes in a nuclear environment is the 
collective nuclear unintegrated glue per unit area in the 
impact parameter space. It is not a single function which
can be defined for the whole nucleus, for all hard processes
of practical interest the description of the initial and final
interactions inevitably calls upon a collective
glue for different slices of the nucleus. Furthermore, such
a collective glue must
be regarded as a density matrix in the space of color
representation, i.e., it changes from one reaction to another
depending on color properties of the relevant pQCD subprocess
\cite{Nonlinear,SingleJet}. One can trace the origin of
these variations to the color-representation dependence 
of the color-dipole cross sections emerging
in the description of these reactions.  

In the treatment of the nuclear structure function
$F_{2A}(x,Q^2)$ and of the quark-antiquark dijets in DIS off nuclei 
it is
advisable to use the collective glue  $\phi(\bb,x,\bkappa)$
defined in terms of the amplitude of coherent diffractive 
quark-antiquark
dijet production \cite{NSSdijet,NSSdijetJETPLett,Nonlinear,NonlinearJETPLett} 
\bea 
1-\exp\left[-{1\over 2}\sigma_{q\bar{q}}(x,\br)T(\bb)\right]
\equiv \int d^2\bkappa
\phi(\bb,x,\bkappa) \Big[1 - \exp(i\bkappa\br) \Big] \, .
\label{eq:D.1} 
\eea
Here \cite{NZ94,NZglue}
\bea 
\sigma_{q\bar{q}}(x,\br) &=& \int d^2\bkappa f (x,\bkappa)
[1-\exp(i\bkappa\br)]\,,
 \label{eq:D.2} 
\eea
where
\bea
f (x,\bkappa)&=& {4\pi\alpha_S(r)\over N_c}
\cdot {1\over \kappa^4} \cdot {\cal
F}(x,\kappa^2)
\label{eq:D.3} 
\eea
and
\beq
{\cal F}(x,\kappa^2) = {\partial G(x, \kappa^2) \over \partial \log \kappa^2}
 \label{eq:D.4}
\eeq 
is  the unintegrated gluon density in the
target nucleon. 

The so-defined collective nuclear glue admits a nice probabilistic
expansion
\bea
\phi(\bb,x,\bkappa)= \sum_{j=1} w_{q\bar{q},j}\Big(\nu_A(\bb)\Big) {f^{(j)}(\bkappa)
\over \sigma_{q\bar{q},0}^j}.
\label{eq:D.5}
\eea
Here
\bea
w_{q\bar{q},j}\Big(\nu_A(\bb)\Big) ={1\over j!} \left[{1\over 2}\nu_A(\bb)\right]^j
\exp\left[-{1\over 2}\nu_A(\bb)\right]
\label{eq:D.6}
\eea
is a probability to find $j$ spatially overlapping nucleons in the
Lorentz-contracted ultrarelativistic nucleus, where  
\beq
\nu_A(\bb)={1\over 2}\sigma_{q\bar{q},0}  T(\bb),
\label{eq:D.7}
\eeq
is the thickness of the nucleus
in terms of the number of absorption lengths for large dipoles,
and we introduced an auxiliary  infrared quantity -- a dipole cross 
section for large quark-antiquark dipoles:
\beq
\sigma_{0,q\bar{q}}= \int d^2\bkappa f(\bkappa) \, .
\label{eq:D.8} 
\eeq
The properly defined $j$-fold convolutions, 
\bea
&&{f^{(j)}(\bkappa)\over \sigma_{q\bar{q},0}^j} = 
\int d^2\bkappa_1 {f^{(j-1)}(\bkappa-\bkappa_1)\over \sigma_{q\bar{q},0}^{j-1}}
\cdot { f(\bkappa_1)\over \sigma_{q\bar{q},0}}, \nonumber\\
&& f^{(0)}(\bkappa) =\delta(\bkappa), \nonumber\\
&&\int d^2\bkappa {f^{(j)}(\bkappa)\over \sigma_{q\bar{q},0}^j} =1,
\label{eq:D.9}
\eea
describe the collective unintegrated glue of $j$ spatially overlapping nucleons
in a Lorentz-contracted nucleus. They do not change from one reaction to another,
the variations from $\phi_g(\bb,\bkappa)$ to $\phi_g(\bb,\bkappa)$ to 
$\phi_{gg}(\bb,\bkappa)$ are fully described by the color-representation 
dependence of the overlap probabilities:
\bea
&&w_{g,j}\Big(\nu_A(\bb)\Big)= w_{q\bar{q},j}\Big({C_A\over 2C_F}\nu_A(\bb)\Big),\nonumber\\
&&w_{gg,j}\Big(\nu_A(\bb)\Big)=w_{q\bar{q},j}\Big({C_A\over C_F}\nu_A(\bb)\Big).
\eea

\end{document}